\documentclass[12pt]{article}

\usepackage{cite}
\usepackage{float} 
\usepackage{graphicx}
\usepackage{amsfonts}
\usepackage{comment}
\usepackage{amsmath}
\usepackage{longtable}

\input{epsf}

\advance \topmargin by -\headheight
\advance \topmargin by -\headsep

\evensidemargin \oddsidemargin
\begin{document}

\topmargin -.6in

\def\rh{{\hat \rho}}
\def\alie{{\hat{\cal G}}}
\newcommand{\sect}[1]{\setcounter{equation}{0}\section{#1}}
\renewcommand{\theequation}{\thesection.\arabic{equation}}

\def\rf#1{(\ref{eq:#1})}
\def\lab#1{\label{eq:#1}}
\def\nonu{\nonumber}
\def\br{\begin{eqnarray}}
\def\er{\end{eqnarray}}
\def\be{\begin{equation}}
\def\ee{\end{equation}}
\def\eq{\!\!\!\! &=& \!\!\!\! }
\def\foot#1{\footnotemark\footnotetext{#1}}
\def\lb{\lbrack}
\def\rb{\rbrack}
\def\llangle{\left\langle}
\def\rrangle{\right\rangle}
\def\blangle{\Bigl\langle}
\def\brangle{\Bigr\rangle}
\def\llbrack{\left\lbrack}
\def\rrbrack{\right\rbrack}
\def\lcurl{\left\{}
\def\rcurl{\right\}}
\def\({\left(}
\def\){\right)}
\newcommand{\nit}{\noindent}
\newcommand{\ct}[1]{\cite{#1}}
\newcommand{\bi}[1]{\bibitem{#1}}
\def\lskip{\vskip\baselineskip\vskip-\parskip\noindent}
\relax

\def\tr{\mathop{\rm Tr}}
\def\Tr{\mathop{\rm Tr}}
\def\trace{\widehat{\rm Tr}}
\def\v{\vert}
\def\bv{\bigm\vert}
\def\Bgv{\;\Bigg\vert}
\def\bgv{\bigg\vert}
\newcommand\partder[2]{{{\partial {#1}}\over{\partial {#2}}}}
\newcommand\funcder[2]{{{\delta {#1}}\over{\delta {#2}}}}
\newcommand\Bil[2]{\Bigl\langle {#1} \Bigg\vert {#2} \Bigr\rangle}  
\newcommand\bil[2]{\left\langle {#1} \bigg\vert {#2} \right\rangle} 
\newcommand\me[2]{\left\langle {#1}\bv {#2} \right\rangle} 
\newcommand\sbr[2]{\left\lbrack\,{#1}\, ,\,{#2}\,\right\rbrack}
\newcommand\pbr[2]{\{\,{#1}\, ,\,{#2}\,\}}
\newcommand\pbbr[2]{\lcurl\,{#1}\, ,\,{#2}\,\rcurl}

\def\ket#1{\mid {#1} \rangle}
\def\bra#1{\langle {#1} \mid}
\newcommand{\braket}[2]{\langle {#1} \mid {#2}\rangle}
%
\def\a{\alpha}
\def\at{{\tilde A}^R}
\def\atc{{\tilde {\cal A}}^R}
\def\atcm#1{{\tilde {\cal A}}^{(R,#1)}}
\def\b{\beta}
\def\dc{{\cal D}}
\def\d{\delta}
\def\D{\Delta}
\def\eps{\epsilon}
\def\vareps{\varepsilon}
\def\g{\gamma}
\def\G{\Gamma}
\def\grad{\nabla}
\def\h{{1\over 2}}
\def\l{\lambda}
\def\L{\Lambda}
\def\m{\mu}
\def\n{\nu}
\def\o{\over}
\def\om{\omega}
\def\O{\Omega}
\def\p{\phi}
\def\P{\Phi}
\def\pa{\partial}
\def\pr{\prime}
\def\pt{{\tilde \Phi}}
\def\qs{Q_{\bf s}}
\def\ra{\rightarrow}
\def\s{\sigma}
\def\S{\Sigma}
\def\t{\tau}
\def\th{\theta}
\def\Th{\Theta}
\def\tpp{\Theta_{+}}
\def\tmm{\Theta_{-}}
\def\tpg{\Theta_{+}^{>}}
\def\tms{\Theta_{-}^{<}}
\def\tp0{\Theta_{+}^{(0)}}
\def\tm0{\Theta_{-}^{(0)}}
\def\ti{\tilde}
\def\wti{\widetilde}
\def\jc{J^C}
\def\bj{{\bar J}}
\def\sj{{\jmath}}
\def\bsj{{\bar \jmath}}
\def\bp{{\bar \p}}
\def\vp{\varphi}
\def\ve{\varepsilon}
\def\vt{{\tilde \varphi}}
\def\faa{Fa\'a di Bruno~}
\def\ca{{\cal A}}
\def\cb{{\cal B}}
\def\ce{{\cal E}}
\def\cg{{\cal G}}
\def\cgh{{\hat {\cal G}}}
\def\ch{{\cal H}}
\def\chh{{\hat {\cal H}}}
\def\cl{{\cal L}}
\def\cm{{\cal M}}
\def\cn{{\cal N}}
\def\u2{\mid u\mid^2}
\newcommand\sumi[1]{\sum_{#1}^{\infty}}   
\newcommand\fourmat[4]{\left(\begin{array}{cc}  
{#1} & {#2} \\ {#3} & {#4} \end{array} \right)}

%
\def\lie{{\cal G}}
\def\kmlie{{\hat{\cal G}}}
\def\dlie{{\cal G}^{\ast}}
\def\elie{{\widetilde \lie}}
\def\edlie{{\elie}^{\ast}}
\def\hlie{{\cal H}}
\def\flie{{\cal F}}
\def\wlie{{\widetilde \lie}}
\def\f#1#2#3 {f^{#1#2}_{#3}}
\def\winf{{\sf w_\infty}}
\def\win1{{\sf w_{1+\infty}}}
\def\hwinf{{\sf {\hat w}_{\infty}}}
\def\Winf{{\sf W_\infty}}
\def\Win1{{\sf W_{1+\infty}}}
\def\hWinf{{\sf {\hat W}_{\infty}}}
\def\Rm#1#2{r(\vec{#1},\vec{#2})}          
\def\OR#1{{\cal O}(R_{#1})}           
\def\ORti{{\cal O}({\widetilde R})}           
\def\AdR#1{Ad_{R_{#1}}}              
\def\dAdR#1{Ad_{R_{#1}^{\ast}}}      
\def\adR#1{ad_{R_{#1}^{\ast}}}       
\def\KP{${\rm \, KP\,}$}                 
\def\KPl{${\rm \,KP}_{\ell}\,$}         
\def\KPo{${\rm \,KP}_{\ell = 0}\,$}         
\def\mKPa{${\rm \,KP}_{\ell = 1}\,$}    
\def\mKPb{${\rm \,KP}_{\ell = 2}\,$}    
%
\def\rlx{\relax\leavevmode}
\def\inbar{\vrule height1.5ex width.4pt depth0pt}
\def\IZ{\rlx\hbox{\sf Z\kern-.4em Z}}
\def\IR{\rlx\hbox{\rm I\kern-.18em R}}
\def\IC{\rlx\hbox{\,$\inbar\kern-.3em{\rm C}$}}
\def\IN{\rlx\hbox{\rm I\kern-.18em N}}
\def\IO{\rlx\hbox{\,$\inbar\kern-.3em{\rm O}$}}
\def\IP{\rlx\hbox{\rm I\kern-.18em P}}
\def\IQ{\rlx\hbox{\,$\inbar\kern-.3em{\rm Q}$}}
\def\IF{\rlx\hbox{\rm I\kern-.18em F}}
\def\IG{\rlx\hbox{\,$\inbar\kern-.3em{\rm G}$}}
\def\IH{\rlx\hbox{\rm I\kern-.18em H}}
\def\II{\rlx\hbox{\rm I\kern-.18em I}}
\def\IK{\rlx\hbox{\rm I\kern-.18em K}}
\def\IL{\rlx\hbox{\rm I\kern-.18em L}}
\def\one{\hbox{{1}\kern-.25em\hbox{l}}}
\def\0#1{\relax\ifmmode\mathaccent"7017{#1}%
B        \else\accent23#1\relax\fi}
\def\omz{\0 \omega}
%
\def\ltimes{\mathrel{\vrule height1ex}\joinrel\mathrel\times}
\def\rtimes{\mathrel\times\joinrel\mathrel{\vrule height1ex}}
%
\def\mark{\noindent{\bf Remark.}\quad}
\def\prop{\noindent{\bf Proposition.}\quad}
\def\theor{\noindent{\bf Theorem.}\quad}
\def\name{\noindent{\bf Definition.}\quad}
\def\exam{\noindent{\bf Example.}\quad}
\def\proof{\noindent{\bf Proof.}\quad}

%
%
\def\PRL#1#2#3{{\sl Phys. Rev. Lett.} {\bf#1} (#2) #3}
\def\NPB#1#2#3{{\sl Nucl. Phys.} {\bf B#1} (#2) #3}
\def\NPBFS#1#2#3#4{{\sl Nucl. Phys.} {\bf B#2} [FS#1] (#3) #4}
\def\CMP#1#2#3{{\sl Commun. Math. Phys.} {\bf #1} (#2) #3}
\def\PRD#1#2#3{{\sl Phys. Rev.} {\bf D#1} (#2) #3}
\def\PLA#1#2#3{{\sl Phys. Lett.} {\bf #1A} (#2) #3}
\def\PLB#1#2#3{{\sl Phys. Lett.} {\bf #1B} (#2) #3}
\def\JMP#1#2#3{{\sl J. Math. Phys.} {\bf #1} (#2) #3}
\def\PTP#1#2#3{{\sl Prog. Theor. Phys.} {\bf #1} (#2) #3}
\def\SPTP#1#2#3{{\sl Suppl. Prog. Theor. Phys.} {\bf #1} (#2) #3}
\def\AoP#1#2#3{{\sl Ann. of Phys.} {\bf #1} (#2) #3}
\def\PNAS#1#2#3{{\sl Proc. Natl. Acad. Sci. USA} {\bf #1} (#2) #3}
\def\RMP#1#2#3{{\sl Rev. Mod. Phys.} {\bf #1} (#2) #3}
\def\PR#1#2#3{{\sl Phys. Reports} {\bf #1} (#2) #3}
\def\AoM#1#2#3{{\sl Ann. of Math.} {\bf #1} (#2) #3}
\def\UMN#1#2#3{{\sl Usp. Mat. Nauk} {\bf #1} (#2) #3}
\def\FAP#1#2#3{{\sl Funkt. Anal. Prilozheniya} {\bf #1} (#2) #3}
\def\FAaIA#1#2#3{{\sl Functional Analysis and Its Application} {\bf #1} (#2)
#3}
\def\BAMS#1#2#3{{\sl Bull. Am. Math. Soc.} {\bf #1} (#2) #3}
\def\TAMS#1#2#3{{\sl Trans. Am. Math. Soc.} {\bf #1} (#2) #3}
\def\InvM#1#2#3{{\sl Invent. Math.} {\bf #1} (#2) #3}
\def\LMP#1#2#3{{\sl Letters in Math. Phys.} {\bf #1} (#2) #3}
\def\IJMPA#1#2#3{{\sl Int. J. Mod. Phys.} {\bf A#1} (#2) #3}
\def\AdM#1#2#3{{\sl Advances in Math.} {\bf #1} (#2) #3}
\def\RMaP#1#2#3{{\sl Reports on Math. Phys.} {\bf #1} (#2) #3}
\def\IJM#1#2#3{{\sl Ill. J. Math.} {\bf #1} (#2) #3}
\def\APP#1#2#3{{\sl Acta Phys. Polon.} {\bf #1} (#2) #3}
\def\TMP#1#2#3{{\sl Theor. Mat. Phys.} {\bf #1} (#2) #3}
\def\JPA#1#2#3{{\sl J. Physics} {\bf A#1} (#2) #3}
\def\JSM#1#2#3{{\sl J. Soviet Math.} {\bf #1} (#2) #3}
\def\MPLA#1#2#3{{\sl Mod. Phys. Lett.} {\bf A#1} (#2) #3}
\def\JETP#1#2#3{{\sl Sov. Phys. JETP} {\bf #1} (#2) #3}
\def\JETPL#1#2#3{{\sl  Sov. Phys. JETP Lett.} {\bf #1} (#2) #3}
\def\PHSA#1#2#3{{\sl Physica} {\bf A#1} (#2) #3}
\def\PHSD#1#2#3{{\sl Physica} {\bf D#1} (#2) #3}
\def\PJA#1#2#3{{\sl Proc. Japan. Acad} {\bf #1A} (#2) #3}
\def\JPSJ#1#2#3{{\sl J. Phys. Soc. Japan} {\bf #1} (#2) #3}
\newcommand{\map}{\mathcal{P}}
\def\tih{\tilde{h}}


\begin{titlepage}
\vspace*{-1cm}

\vskip 3cm

\vspace{.2in}
\begin{center}
{\large\bf Self-duality and the Holomorphic Ansatz in Generalized BPS Skyrme Model}
\end{center}

\vspace{.5cm}

\begin{center}
L. A. Ferreira$^{\dagger,}$\footnote{laf@ifsc.usp.br} and L. R. Livramento$^{\dagger, }$\footnote{livramento@usp.br}

\vspace{.3 in}
\small

\par \vskip .2in \noindent
$^{\dagger}$Instituto de F\'\i sica de S\~ao Carlos; IFSC/USP;\\
Universidade de S\~ao Paulo, USP  \\ 
Caixa Postal 369, CEP 13560-970, S\~ao Carlos-SP, Brazil\\

\normalsize
\end{center}

\vspace{.5in}

\begin{abstract}

We propose a generalization of the BPS Skyrme model \cite{laf2017} for simple compact Lie groups $G$ that leads to Hermitian symmetric spaces. In such a theory, the Skyrme field takes its values in $G$, while the remaining fields correspond to the entries of a symmetric, positive, and invertible $\dim G \times \dim G$-dimensional matrix $h$. We also use the holomorphic map ansatz between $S^2 \rightarrow G/H \otimes U(1)$ proposed in \cite{Ferreira:2024ivq} to study the self-dual sector of the theory, which generalizes the holomorphic ansatz between $S^2 \rightarrow CP^N$ proposed in \cite{Ioannidou:1998zg}. This ansatz is constructed using the fact that stable harmonic maps of the two $S^2$ spheres for compact Hermitian symmetric spaces are holomorphic or anti-holomorphic \cite{harmonicmaps}.  Apart from some special cases, the self-duality equations do not fix the matrix $h$ entirely in terms of the Skyrme field, which is completely free, as it happens in the original self-dual Skyrme model for $G=SU(2)$. In general, the freedom of the $h$ fields tend to grow with the dimension of $G$.  The holomorphic ansatz enable us to construct an infinite number of exact self-dual Skyrmions for each integer value of the topological charge and for each value of $N \geq 1$, in case of the $CP^N$, and for each values of $p,\,q\geq 1$ in case of $SU\(p+q\)/SU\(p\)\otimes SU\(q\)\otimes U\(1\)$.

\end{abstract} 
\end{titlepage}

\section{Introduction}
\label{sec:intro}
\setcounter{equation}{0}

The study of self-duality has shed light on the complex behavior of topological solutions in a wide variety of classical nonlinear field theories. The topological solitons are classified by a homotopic invariant quantity, the so-called topological charge, and self-duality can greatly facilitate the task of obtaining the topological solutions corresponding to the global energy minimizer \cite{genbps}. This plays a fundamental role in the study of kinks and instantons in $(1+1)$ dimensions \cite{rajaraman1982solitons, mantonbook, Shnir:book1, genbps}, vortex solutions in the abelian Chem-Simons theory in $(2+1)$ dimensions \cite{Jackiw:1990aw}, self-dual Skyrmions in $(3+1)$ dimensions \cite{laf2017, luiz1} and in some non-abelian gauge theories in $(3+1)$ dimensions, as the Yang-Mills-Higgs System \cite{Ferreira:2021uhk}. 

The self-duality usually appears in models that possess two main ingredients. First, the static energy density of the model must be the sum of the squares of two objects $A_\alpha$ and $\tilde{A}_\alpha$ that depend on the fields and their first-order space-time derivatives only, where the nature of the fields and the $\alpha$ index depends on each theory. Second, the topological charge density must be proportional to the contraction of such objects. It follows that the so-called self-duality equations $A_\alpha=\pm \,\tilde{A}_{\alpha}$ imply second-order differential Euler-Lagrange equations and also correspond to the global minimizer of the static energy, for each value of the topological charge ($Q$). The set of topological solutions of the self-duality equations is called the self-dual sector, which can be empty for some models, as is the case with the standard Skyrme model, as demonstrated in \cite{mantonruback, derek}.

The standard Skyrme model is an effective classical field theory for the triplet of pions in $(3+1)$ dimensions in the low-energy regime \cite{skyrme1,skyrme2,mantonbook,mantonruback,adkins}. The model is defined in terms of the $SU(2)$ Skyrme field $U$, which includes the three pion fields and is a map between two three-spheres. Its standard version contains only two terms in its action, one quadratic and the other quartic in the space-time derivatives. The quartic term, or any other higher-order kinetic term, is essential to stabilize the Skyrmions under Derrick's scale argument \cite{derrick}. This is still true even if any positive definite potential defined in terms of the Skyrme field is added since this $SU(2)$ field is scale invariant. Such a theory possesses a large number of modifications, some allowing the construction of electrically charged multi-Skyrmions. This is the case of the gauged version of the Skyrme model obtained by gauging the $U(1)$ subgroup of the $SU(2)$ global symmetry, associated with the generator of its Cartan subalgebra \cite{Callan:1983nx, Piette:1997ny, Radu:2005jp, Livramento:2023keg, Livramento:2023tmm}.

There are some modifications to the standard Skyrme model that lead to a non-empty self-dual sector. Notably, one of these modifications, the so-called BPS Skyrme model \cite{laf2017}, can be directly derived from integral representations of the topological charge associated with the Skyrme field using ideas of self-duality seen in \cite{genbps}. Such an approach spontaneously includes six extra fields corresponding to the entries of a symmetric, positive, and invertible $3\times 3$ matrix $h$. The matrix $h$ and its inverse appear contracted respectively to quadratic and quartic terms in the space-time derivatives associated with the Skyrme field, i.e., the model is defined by
\be S_{BPS}=\int d^4x\left[\frac{m_0^2}{2}h_{ab}R_\mu^a R^{b,\mu}-\frac{1}{4\,e_0^2}h_{ab}^{-1}H_{\mu\nu}^aH^{b,\mu\nu}\right]\lab{skyrmebps} \ee
where $m_0$ is a coupling constant with dimension of mass, and $e_0$ is a dimensionless coupling constant. In addition,  $R^a_{\mu}=i\, \trace\(\partial_{\mu}U\,U^{\dagger}\,T_a\)$, and $H^a_{\mu\nu}=\ve_{abc}\,R_{\mu}^b\,R_{\nu}^c$,  with $T_a$, $a=1,2,3$, being the basis of the $SU(2)$ Lie algebra satisfying $\sbr{T_a}{T_b}=i\,\ve_{abc}\,T_c $, and $\trace\(T_a\,T_b\)=\delta_{ab}$. The standard Skyrme model is recovered by imposing $h=\one$.

It was demonstred in \cite{luiz1} that self-duality equations of the BPS Skyrme model can be used to algebraically determine entierly the $h$ matrix in terms of the matrix $\tau_{ab}\equiv R_i^a\,R_i^b$, with $a,\,b=1,\,2,\,3$, in all regions where $\tau$ is non-singular, while the Skyrme field still completely free. This $SU(2)$ field is still completely free even at the points where $\tau$ is singular, but in this case some of the components of the matrix $h$ are also free. The reason that leads to this freedom can be traced to the fact that the nine static Euler-Lagrange equations for the fields $h$ and $U$ are not all independent. In fact, the equations for the $U$-field can be derived from the equations of the $h$-fields when $\tau$ is non-singular. The freedom of the Skyrme field leads to an infinite number of exact topological solutions to each value of the topological charge.

All the BPS solutions of the model \rf{skyrmebps} are scale independent due to the conformal invariance of the model in three spacial dimensions. This freedom of the shapes of topological solitons can improve the scope of physical application of the theory, especially if some extra term is added breaking the scale independence and selecting some specific form. By example, the scale dependence and the radial multi-solitons configurations that live in the self-dual sector of the theory \rf{skyrmebps} are essential in one of its extensions, the False Vacuum Skyrme model \cite{luiz:false}.

Self-duality can also play a fundamental role in models that are extensions of BPS theories, where the total static energy contains extra terms, even in non-perturbative approaches. On the one hand, self-duality can inspire the construction of ansätze in quasi-self-dual models, where the extra terms weakly break the self-duality equations \cite{quasi-self-dual}. On the other hand, there are models that contain extra terms that do not break any of the self-duality equations, such as the False Vacuum Skyrme model \cite{luiz:false}. This is a powerful modification of the Skyrme model that leads to excellent classical results for the binding energy and radius of the nuclei. In fact, the results are such that for a list containing  $256$ nuclei with mass number $A\geq 12$, the root-mean-square deviation of the binding energy per nucleon and the root-mean-square radius, which are respectively of the order of $0.05\, MeV$ and  $0.04\, fm$, are of the same order as excellent fits based on phenomenological approaches.

The magic of the False vacuum BSP Skyrme model is that the BSP Skyrme term gives a massive contribution $E_1$ to the total nuclear mass, but the binding energy comes just from the extra terms, despite given a lower order contribution $E_2$  to the total mass. The BPS model is extended trough the introduction of kinetic and potential terms for the baryonic density, which depends only on the Skyrme field, and a topological term that approximately reproduces the Coulomb interaction. The $h$-fields still being determined through the self-duality equations, since the additional terms do not depend on such fields. Curiously, Coleman's false vacuum argument \cite{coleman1, coleman2, colemannew} shows that the global minimizer of such a theory must have radial symmetry. This mathematical result reduces the three-dimensional static Euler–Lagrange equations to a single radial second order differential equation for a fractional power of the baryonic density. However, exploring the nature of $h$ fields for generalizations of the BPS Skyrme model can help reveal the physical interpretation of such fields and if $h$ can still be entirely determined in terms of the Skyrme fields for Lie groups other than $G=SU(2)$.

An important mathematical result that motivates us to generalize the BPS Skyrme model shows that for a simple compact Lie group $G$, it follows that $\pi_3(G)=\IZ$, where $\pi_3(G)$ is the homotopy group of the mapping of $G$ into a $3$-sphere. The topological charge associated with this map admits integral representation similar to the $G=SU(2)$ case. Let us consider now the cases where $G$ lead to a Hermitian symmetric space $G/H\otimes U(1)$, where little group $H\otimes U(1)$ is a subgroup of $G$. Some examples of Lie groups that lead to a Hermitian symmetric space are $G=A_r,\, B_r,\, C_r,\,D_r,\,E_6,\,E_7$, and some counterexamples include the Lie Groups $G=E_8,\,F_4,\,G_2$. 

Another important mathematical result that can shed light on how we can study the self-duality in such model was derived by Eells and Lemaire \cite{harmonicmaps}. It states that stable harmonic maps $X$ from the two-sphere $S^2$ to compact Hermitian symmetric spaces are holomorphic or anti-holomorphic. This laid the  foundation for constructing the ansatz holomorphic map ansatz between $S^2 \rightarrow G/H\otimes U(1)$ proposed in \cite{Ferreira:2024ivq} by L. A. Ferreira and L. R. Livramento. Despite this ansatz having some conditions for applicability that depend on the representation of $G$, it is major generalization of the holomorphic ansatz  between $S^2 \rightarrow CP^N$  proposed in \cite{Ioannidou:1998zg} to the $CP^N$.

The main idea of this work is to first construct a generalized BPS Skyrme model with the Skyrme fields mapping the physical space to a simple compact Lie group $G$ that leads to the Hermitian symmetric space $G/H\otimes U(1)$. In this case, as the indices of the rows and columns of the matrix $h$ are contracted with each index of the generators of the Lie algebra ${\cal G}$ associated with $G$, in such a theory it becomes a $\dim { \cal G}\times \dim {\cal G}$ dimensional symmetric, invertible and positive matrix. Therefore, the $h$ matrix and the Skyrme fields can be written in terms of $\dim {\cal G}\,(\dim {\cal G}+1)/2$ and $\dim {\cal G}$ independents fields, respectively.

Our second objective in this paper is to study the self-dual sector of such a theory through the holomorphic ansatz between $S^2 \rightarrow G/H\otimes U(1)$ proposed in \cite{Ferreira:2024ivq}. In particular, we want to determine whether the matrix $h$ can still be entirely determined in terms of the Skyrme fields in the generalized BPS Skyrme model, similar to what happens in the case $G=SU(2)$, and whether $U$ is still completely free. Despite the fact that the number of self-duality equations is, in principle, equal to the number of independent fields of the theory, the full determination of all the fields by the self-duality equations is not expected, since this does not happen even for the $G=SU(2)$ case, as discussed above. This ansatz can drastically simplify the self-duality equations, aiding in our investigation of the self-dual sector and in the construction of exact BPS topological solutions.

A powerful holomorphic ansatz for the standard Skyrme model was constructed for the $G=SU(2)$ case by Houghton et al. in \cite{rational1} using harmonic maps from $S^3 \rightarrow S^3$ . It is based on the rational map, which is a holomorphic function from $S^2\rightarrow S^2$ \cite{rational1, rational2, mantonbook, Shnir:book1}. Although solving all Euler-Lagrange equations only for $Q=\pm 1$, such ansatz leads to a quite good approximation of the true topological solitons corresponding to the global energy minimizer. However, in the BPS Skyrme model \rf{skyrmebps}, as the Skyrme field is completely free inside the self-dual sector, the rational map leads to an infinite number of exact solutions for each value of $Q$ \cite{luiz1}. 

In interpreting the Skyrme model as a low-energy effective field theory of QCD  in the limit where the number of colors is large, the number $N+1$ of the Lie group $SU(N+1)$ where the Skyrme field takes its values correspond to the number of light quark flavors. The holomorphic ansatz proposed by Houghton et al. in \cite{rational1} for $G=SU(2)$ was generalized in \cite{Ioannidou:1998zg} for $G=SU(N+1)$ using harmonic maps from $S^2$ to $CP^N \cong SU(N+1)/SU(N)\otimes U(1)$. As the others ansatzë used to construct multi-Skyrmions for some values of $N$ of the $G=SU(N+1)$ case, the goal of this ansatz is just give some aproximation of the global energy minimizers. However, in general the energies obtained through such ansatz are marginally higher than the ones obtained through $SU(2)$ embeddings. 

Through the holomorphic ansatz we can construct an infinite number of exact topological solutions for each values of $Q$ and $N$ of the $CP^N$ space. This set of self-dual solutions even includes field configurations based on the standard rational map ansatz. We also obtain self-dual solutions within this ansatz for the Hermitian symmetric space $SU\(p+q\)/SU\(p\)\otimes SU\(q\)\otimes U\(1\)$, which generalizes our results obtained for the $CP^N$. In this case, we also provide explicit solutions for any value of the topological charge. This study has the potential to reveal the nature of $h$ fields in such models and how the role of self-duality manifests in determining the $h$ and $U$ fields.

The paper is organized as follows. In Section \ref{sec:model}, we construct the generalized BPS Skyrme model. Additionally, we obtain the self-dual equations and the expression for the topological charge and energy within the self-dual sector. In this section we also discuss an important symmetry of the model under the composition of parity and target space parity transformations. In Section \ref{sec:equations}, we derive the Euler-Lagrange equations and demonstrate how they can be solved using the self-dual field configurations. In section \ref{sec:holomorphic} we construct our holomorphic ansatz between $S^2 \rightarrow G/H\otimes U(1)$. We also obtain the explicit general form of the self-duality equations using the structure of the Hermitian symmetric space and its our holomorphic ansatz. In sections \ref{sec:su2} and \ref{sec:sun} we study our holomorphic ansatz in the $G=SU(2)$ and $G=SU(N+1)$ case, respectively. The case $SU(2)$ is done separately due to its peculiar structure, and in both cases we obtain exact multi-BPS Skyrmions for all integer values of the topological charge. In section \ref{Spq} we obtain the self-dual equations inside the holomorphic ansatz for the Hermitian symmetric space  $SU\(p+q\)/SU\(p\)\otimes SU\(q\)\otimes U\(1\)$, and construct particular self-dual solutions for all integer values of the topological charge. In the section \ref{sec:conclusion} we present our final considerations.

\section{The model and its construction}
\label{sec:model}
\setcounter{equation}{0}

Consider a simple compact Lie group $G$. It is known that the maps  $S^3 \rightarrow G$ are classified by the integers since 
\be
\pi_3\(G\)= \IZ
\ee
The topological charge associated to such homotopy group is given by
\be
 Q= \frac{i}{48\,\pi^2\,\kappa}\int d^3x\; \ve_{ijk}\,{\rm Tr}\(R_i\,R_j\,R_k\)
 \lab{topcharge}
 \ee
 where
 \be
R_{\mu} \equiv i\,\partial_{\mu}U\,U^{-1}\equiv R^a_{\mu}\,T_a \qquad\qquad\qquad 
\lab{rdef}
\ee
with $U$ being an element of the group $G$, and $T_a$, $a=1,2,3,\ldots {\rm dim}\,G$,  being the generators of the corresponding compact simple Lie algebra
\be
\sbr{T_a}{T_b}=i\,f_{abc}\,T_c 
\ee
and where we work with an orthogonal basis, i.e.
\be
{\rm Tr}\(T_a\,T_b\)=\kappa\, \delta_{ab} \lab{kappadef}
\ee
with $\kappa$ depending upon the representation where the trace is taken. We shall use a normalised trace define as 
\be
\trace\(T_a\,T_b\)\equiv \frac{1}{\kappa}\,{\rm Tr}\(T_a\,T_b\)= \delta_{ab} \lab{normalizedtr}
\ee

The quantities $R_{\mu}$ introduced in \rf{rdef} satisfies by construction the Maurer-Cartan equation
\be
\partial_{\mu}R_{\nu}-\partial_{\nu}R_{\mu}+i\,\sbr{R_{\mu}}{R_{\nu}}=0
\lab{maurercartan} 
\ee
which allow us to split the topological charge \rf{topcharge} as
\be
Q= \frac{1}{48\,\pi^2}\int d^3x\; {\cal A}_i^a\, {\widetilde{\cal A}}_i^a \lab{Qtop0}
\ee
with 
 \be
{\cal A}_i^a\equiv R_i^b\, k_{ba} \; ; \qquad\qquad\qquad \qquad 
{\widetilde{\cal A}}_i^a\equiv \frac{i}{2}\,k^{-1}_{ab}\,\ve_{ijk}\,\trace\(T_b\sbr{R_j}{R_k}\)
\lab{selfdualityeqs}
\ee
where $k_{ab}$ is some invertible matrix. Using the ideas of self-duality seen in \cite{genbps}, through this spliting we can introduce the self-duality equation as
\be
\lambda\,{\cal A}_i^a={\widetilde{\cal A}}_i^a \qquad\qquad {\rm with} \qquad \lambda=\pm m\,e \lab{self0}
\ee
or
\be
\lambda\,R_i^b\,h_{ba}= \frac{i}{2}\,\ve_{ijk}\,\trace\(T_a\sbr{R_j}{R_k}\)
\lab{selfdualityeqs2}
\ee
where we have introduced a $\dim G \times \dim G$-dimensional matrix 
\be
h_{ab}=\(k\,k^T\)_{ab}= k_{ac}\,k_{bc} \lab{defh0}
\ee
Due the fact the $k$ is invertible and the definition \rf{defh0}, it so follows that the $h$ matrix is invertible, symmetric and positive. The fact that $h$ is positive is less trivial, but consider real vector $v$ and define $u\equiv k^T\,v$, which implies that $v=\vec{0}\Rightarrow u=\vec{0}$. Using the fact that $k$ is invertible, we can write $v=k^{T\,-1}\,u$, and so $u=\vec{0}\Rightarrow v=\vec{0}$. Therefore,  the fact that $k$ is invertible implies that $u=\vec{0}\Leftrightarrow v=\vec{0}$. It so follows that for all non-vanishing real vector $v$ we have $v^T\,h\,v =  \mid k^T v \mid^2  = \mid u \mid^2  >0$, and so $h=k\,k^T$ is a positive matrix.

The solutions of \rf{selfdualityeqs} solve the Euler-Lagrange equations associated to the following static energy functional constructed using ideas of self-duality \cite{genbps}
\br
E&=&\frac{1}{2}\,\int d^3x\left[m^2\, \({\cal A}_i^a\)^2+\frac{1}{e^2}\, \({\widetilde{\cal A}}_i^a\)^2\right]
\lab{skyrmeenergy}\\
&=&\frac{1}{2}\,\int d^3x\left[m^2\,h_{ab}\,R_i^a\,R_i^b-\frac{1}{2\,e^2}\,h^{-1}_{ab}\,\trace\(T_a\sbr{R_j}{R_k}\)\,\trace\(T_b\sbr{R_j}{R_k}\)\right]
\nonumber
\er
which is the static energy of a generalized Skyrme model. The action associated to energy \rf{skyrmeenergy} that defines the generalized BPS Skyrme model is so given by
\br
S&=&\frac{1}{2}\,\int d^4x\left[m^2\,h_{ab}\,R_{\mu}^a\,R^{b\,,\,\mu}+\frac{1}{2\,e^2}\,h^{-1}_{ab}\,\trace\(T_a\sbr{R_{\mu}}{R_{\nu}}\)\,\trace\(T_b\sbr{R^{\mu}}{R^{\nu}}\)\right]
\nonumber\\
&=&\int d^4x\left[\frac{m^2}{2}\,h_{ab}\,R_{\mu}^a\,R^{b\,,\,\mu}-\frac{1}{4\,e^2}\,h^{-1}_{ab}\,H_{\mu\nu}^a\,H^{b\,,\,\mu\nu}\right] \lab{action0}
\er
where we have defined (see \rf{maurercartan})
\be
H_{\mu\nu}^a\equiv -i\, \trace\(T_a\sbr{R_{\mu}}{R_{\nu}}\)=\partial_{\mu}R_{\nu}^a-\partial_{\nu}R_{\mu}^a=f_{abc}\,R_{\mu}^b\,R_{\nu}^c \lab{Hdef}
\ee

We can write the energy \rf{skyrmeenergy} as
\br
E&=&\frac{1}{2\,e^2}\,\int d^3x\left[m^2\,e^2\, \({\cal A}_i^a\)^2+ \({\widetilde{\cal A}}_i^a\)^2\right]
\nonumber\\
&=&
\frac{1}{2\,e^2}\,\int d^3x\left[\lambda\, {\cal A}_i^a- {\widetilde{\cal A}}_i^a\right]^2+\frac{\lambda}{e^2}\,\int d^3x\,{\cal A}_i^a\,{\widetilde{\cal A}}_i^a
\nonumber\\
&=&
\frac{1}{2\,e^2}\,\int d^3x\left[\lambda\, {\cal A}_i^a- {\widetilde{\cal A}}_i^a\right]^2+{\rm sign}\,\(\lambda\) \,48\,\pi^2\,\frac{m}{e}\,Q \:\: \geq \:\: {\rm sign}\,\(\lambda\) \,48\,\pi^2\,\frac{m}{e}\,Q \quad \lab{energysd0}
\er
which corresponds to the usual BPS bound. When the self-duality \rf{self0} holds true the topological charge  \rf{Qtop0} can be written as
\be
Q= \pm\frac{m\,e}{48\,\pi^2}\int d^3x\; \({\cal A}_i^a\)^2
\ee
and so $Q$ is positive for the plus sign $\(\lambda>0\)$ and negative otherwise $\(\lambda<0\)$, i.e.
\be {\rm sign}\(Q\,\lambda\)=1\lab{signQ}\ee
Then, using \rf{self0} and \rf{signQ} the energy \rf{energysd0} of the self-dual solutions saturares the BPS bound, also given in \rf{energysd0}, i.e. the energy become
\be
E= 48\,\pi^2\,\frac{m}{e}\,\mid Q\mid \lab{sdenergy1}
\ee
Clearly, as usually, the self-dual energy \rf{sdenergy1} is proportional to the modulus of the topological chage. 

Contracting the self-duality equations \rf{selfdualityeqs2} with $R_i^c$ we get 
\be
\lambda\, \tau_{cb}\,h_{ba}=\sigma_{ca}
\lab{sdconsequence}
\ee
with
\be
\tau_{ab}\equiv R_i^a\,R_i^b
\lab{taudef}
\ee
and
\be
\sigma_{ab}\equiv \frac{i}{2}\,R_i^a\,\ve_{ijk}\,\trace\(T_b\sbr{R_j}{R_k}\)=
-\frac{1}{2}\,\ve_{ijk}\,f_{bcd}R_i^a\,R_j^c\,R_k^d = 
-\frac{1}{2}\,\ve_{ijk}\,R_i^a\,H_{jk}^b 
\lab{sigmadefmatrix}
\ee
Note that the self-dual equations \rf{selfdualityeqs2} are labeled by one spatial index $i=1,\,...,\, 3$ and one algebra index $a=1,\,...,\, \dim G$, while, due to the contraction with $R_i^c$, the equations \rf{sdconsequence} are labeled by two algebraic indices $a,\,c=1,\,...,\, \dim G$. In Section \ref{sec:equations}, we show that the $\dim G \times \dim G$ self-duality equations \rf{sdconsequence} are equivalent to the $\dim G \times 3$ self-duality equations \rf{selfdualityeqs2}.

From \rf{topcharge} and \rf{sigmadefmatrix} the topological charge become 
\br
 Q&=& \frac{i}{96\,\pi^2}\int d^3x\; \ve_{ijk}\,\trace\(R_i\,\sbr{R_j}{R_k}\)
 =-\frac{1}{96\,\pi^2}\int d^3x\;\ve_{ijk}\,f_{abc}\,R_i^a\,R_j^b\,R_k^c
 \nonumber\\
 &=&\frac{1}{48\,\pi^2}\int d^3x\;\sigma_{aa}
 \lab{topcharge2}
 \er
Note that in the particular cases where $\tau$ is invertible we can write $h$ in terms of the $U$-fields only as
\be
h=\frac{1}{\lambda}\, \tau^{-1}\,\sigma
\lab{hsolve}
\ee
So, the self-duality equation is solved for any $U$-field configuration (as long as $\tau$ is invertible at least), and so the $h$-fields are spectators in the sense they adjust themselves to that $U$-configuration to solve the self-duality equations. However, in the case $\tau$ is not invertible, the BPS Skyrmions need to be constructed by solving the self-duality equations \rf{sdconsequence}.

Under space parity $P$ transformations $(t,\,x_i)\rightarrow (t,\,-x_i)$ and under the target space parity $P_U$ transformations $U\rightarrow U^{-1}$, where $U$ can be any element of the target space $G$, we have the same transformations for the quantities $(\tau,\, \sigma,\, h,\,E)\rightarrow (\tau,\, -\sigma,\, -h,\,-E)$. Note that the way that $h$ transforms under $P$ and $P_U$ can be derived from the way that $\sigma$ and $\tau$ transforms using the self-duality equations \rf{sdconsequence}. Clearly, by the space parity transformations $\tau$ is a scalar, while $E$, $\sigma$ and the $h$ fields are pseudo-scalars.  These two set of transformations shows in particular that the energy $E$ is invariant under the composition $P\,P_U$. 

The fact that the $h$ fields gets a minus sign in both transformations $P$ and $P_U$ lead to an important distinction of our theory and the standard Skyrme model, where by definition $h=\one$ does not transforms. The energy of the standard Skyrme model $E_{{\rm Sk}}$ is also invariant under the composition $P\,P_U$, but this comes from the fact $E_{{\rm Sk}}$ is also invariant under both $P$ and $P_U$ transformations separately. 

The $h$ fields plays the same role of the Wess-Zumino term with respect the $P$ and $P_U$ transformations. The Wess-Zumino term is introduced into the Skyrme model in \cite{witten1} to break both invariances $P$ and $P_U$ while preserving the invariance $P\,P_g$. This is fundamental in the interpretation of the Skyrme model as an effective theory in the low energy regime, where just the composition $P\,P_U$ must be a symmetry of the action. In fact, for three flavors the Skyrme field takes its values in the $SU(3)$ Lie group and can be written in terms of an octet formed by pions, Kaons and eta mesons \cite{Holzwarth:1985rb, Weigel:1995cz, Schechter:1999hg, Loiseau:1989ja}. The $P_U$ invariance would forbids, for example, the process $K^+\,K^- \rightarrow \pi^+ \, \pi^- \, \pi^0$, where $K^+$ is the kaon, $K^-$ the anti-kaon, and $(\pi^+ ,\, \pi^- ,\, \pi^0)$ corresponds to the three pions with electrical charges $+e,\,-e,\,0$, respectively, where $e$ is the electric charge of the proton. However, this process can be observed experimentally and is allowed in QCD by the non-abelian anomaly.

\section{The Euler-Lagrange equations}
\label{sec:equations}
\setcounter{equation}{0}
The Euler-Lagrange equations associated to the Skyrme field and the action \rf{action0} are
\be \pa_\mu\left( - \lambda^2 \,h_{ab}\,R^{b,\,\mu} +  f_{cba} \,R^{c}_{\nu} \, h_{bd}^{-1}\, H^{d,\,\mu\nu} \right)- f_{cba}\,\left[-\lambda^2 \, h_{bd}\,R_\mu^d \,R^{c,\,\mu}   + h_{bd}^{-1}\,H^{d,\,\mu\nu}\,\pa_\mu R_\nu^c\right] =0\ee
Its static version is given by
\be \pa_i\left(  \lambda^2 \, h_{ab} \,R_i^b +  f_{cba}\, R_j^c \,  h_{bd}^{-1} \,H_{ij}^d \right)- f_{cba}\, \left[\lambda^2 \, h_{bd}R_i^d \, R_i^c   + h_{bd}^{-1} \, H_{ij}^d\, \pa_i R_j^c\right] =0  \lab{staticu}\ee
The Euler-Lagrange equations associated to the $h_{ab}$ fields and the action \rf{action0} are
\be
\lambda^2\,R_{\mu}^a\,R^{b\,,\,\mu}+\frac{1}{2}\,h^{-1}_{ac}\,h^{-1}_{bd}\,H_{\mu\nu}^c\,H^{d\,,\,\mu\nu}=0
\lab{heleq}
\ee
Let us introduce 
\be
S_i^{(\pm),\, a} \equiv \mid \lambda\mid\, R_{i}^a \pm \frac{1}{2}\,\ve_{ijk}\,h^{-1}_{ac}\,H_{jk}^c \lab{sia} 
\ee
which satisfies by construction
\be S_i^{(+),\, a}\,S_i^{(-),\, b} = B_{ab}+ A_{ab} \lab{sab1}\ee
where 
\be 
B_{ab} \equiv m^2\,e^2\,R_{i}^a\,R_i^{b}-\frac{1}{2}\,h^{-1}_{ac}\,h^{-1}_{bd}\,\,H_{ij}^c\,H^{d}_{ij} \;;\qquad A_{ab} \equiv \mid \lambda \mid \, \left[\(\sigma\,h^{-1}\)_{ab}-\(\sigma\,h^{-1}\)_{ba}\right] \lab{sab2}
\ee
The equation \rf{sab1} splits $S_i^{(+),\, a}\,S_i^{(-),\, b}$ into its symmetric and anti-symmetric parts $B_{ab}$ and $A_{ab}$, respectively. The static version of \rf{heleq} becomes
\be B_{ab}=0 \;\qquad\qquad \Leftrightarrow \qquad \qquad S_i^{(+),\, a}\,S_i^{(-),\, b} = A_{ab} \lab{sss1}\ee
On the other hand, the self-duality equations \rf{selfdualityeqs2} can be written as 
\be {\rm sign}\,\(\lambda\) =\pm 1 \qquad \Rightarrow \qquad S_i^{(\pm),\, a} = 0 \lab{self11} \ee
which implies 
\be S_i^{(+),\, a}\,S_i^{(-),\, b} = 0 \lab{self22} \ee

Let us show that the $\dim G \times \dim G$ self-duality equations \rf{sdconsequence} are equivalent to the $\dim G \times 3$ self-duality equations \rf{selfdualityeqs2}, which can also be written as \rf{self11}. The equation \rf{sdconsequence} is obtained in Section \ref{sec:model} from \rf{selfdualityeqs2}. Now, let us prove that \rf{sdconsequence} implies \rf{selfdualityeqs2}. In particular, the self-duality equations are also solutions of the static Euler-Lagrange equation associated with the $h$ field \rf{sss1}, i.e., $B_{ab}=0$. On the other hand, from \rf{sdconsequence}, we obtain that $\sigma\, h^{-1}$ is symmetric, and due to \rf{sab2}, we have $A_{ab}=0$. Therefore, the r.h.s. of \rf{sab1} vanishes, reducing this equation to \rf{self22}. The definition \rf{sia} leads to
\br
S_i^{(+),\, a}+\,S_i^{(-),\, a} & =& 2 \, \mid \lambda\mid\, R_{i}^a \lab{cond1}\er
Contracting \rf{cond1} with $S_i^{(\pm),\, b} $ and using \rf{self22} we obtain
\br
S_i^{(\pm),\, a}\,S_i^{(\pm),\, b} &=& 2 \, \mid \lambda\mid\, R_{i}^a\,S_i^{(\pm),\, b} =  2\,\mid \lambda\mid\,\left[ \mid \lambda\mid\,\tau_{ab} \mp \(\sigma\,h^{-1}\)_{ab}\right] \nonumber\\
&=& 2\,\mid \lambda\mid\,\left[ \mid \lambda\mid \mp \,\lambda \right]\,\tau_{ab} 
\er
where we use \rf{sdconsequence}, which implies the self-duality equations \rf{self11}, completing the proof. 

Now, let us explicitly show that the Euler-Lagrange equation for the $U$ field \rf{staticu} is implied by the self-duality equations \rf{selfdualityeqs2}. Using \rf{Hdef} we also can write \rf{selfdualityeqs2} as \be \varepsilon_{ijk}\,\pa_j R_k^a=\frac{1}{2}\,\varepsilon_{ijk} \, H_{jk}^a=-\lambda \, R_i^b \, h_{ba}\qquad \Rightarrow \qquad H_{ij}^a=-\lambda\,\varepsilon_{ijk} \, R_k^b \,h_{ba} \lab{hforms}\ee
Consequently, $\pa_i\left( h_{ab}R_i^a \right)=- \lambda^{-1}\varepsilon_{ijk}\pa_i\pa_j  R_k^b=0$. Using this expression, and defining $L_a$ as the l.h.s. of \rf{staticu}, we obtain
\br L_a & \equiv & \pa_i\left( \lambda^2 \, h_{ab} \,R_i^b +  f_{cba}\, R_j^c \,  h_{bd}^{-1} \,H_{ij}^d \right) - f_{cba}\, \left[\lambda^2 \, h_{bd}R_i^d \, R_i^c   + h_{bd}^{-1} \, H_{ij}^d\, \pa_i R_j^c\right]  \nonumber \\
& = & f_{cba}\, R_j^c \, \pa_i\left( h_{bd}^{-1} \,H_{ij}^d \right) - f_{cba}\, \lambda^2 \, h_{bd}R_i^d \, R_i^c   \lab{staticu1}
\er
However, \rf{hforms} implies  $R_j^c \, \pa_i\left( h_{bd}^{-1} \,H_{ij}^d \right)=  -\lambda\,R_j^c \,\varepsilon_{ijk} \,  \pa_i\left( h_{bd}^{-1} \, R_k^l \,h_{ld}\right)=-\lambda\, R_j^c \,\varepsilon_{ijk}  \, \pa_i R_k^b= \lambda\, R_i^c \, \varepsilon_{ijk} \,\pa_j R_k^b= - \lambda^2\, R_i^c \, R_i^d\,h_{db}$. Then, the first and second terms on the l.h.s. of \rf{staticu1} are equal, and we can write it as
\br L_a & = & 2\,f_{cba}\, \lambda\, R_i^c \, \varepsilon_{ijk} \,\pa_j R_k^b = -2\,i\,\lambda\, \trace\(C\,T_a\)  \lab{staticu2}
\er
where we use $f_{cba} = -i\,\trace \(\sbr{T_c}{T_b}\,T_a\)$ and 
\br
C & \equiv & \varepsilon_{ijk} \,\sbr{R_i}{\pa_j R_k} =\varepsilon_{ijk} \,\(\pa_j\sbr{R_i}{R_k} -\sbr{\pa_j R_i}{R_k} \)=-i\,\varepsilon_{ijk} \,\pa_j\(\pa_i R_k-\pa_k R_i\) \nonumber\\
&& -\varepsilon_{ijk} \,\sbr{\pa_j R_i}{R_k} =-\varepsilon_{ijk} \,\sbr{\pa_j R_i}{R_k} =  -\varepsilon_{ijk} \,\sbr{R_i}{\pa_j R_k} = - C 
\er
Therefore, $C=0$ leading due to \rf{staticu2} to $L_a=0$, which corresponds to the static Euler-Lagrange equations for the Skyrme field \rf{staticu}.

In the self-dual sector, the equation \rf{selfdualityeqs2} implies that $\sigma\,h^{-1}$ is symmetric. Therefore, $A_{ab}=0$, and using \rf{self22} we obtain that the self-duality equations \rf{self11} imply the static Euler-Lagrange equations for the $h$ field \rf{sss1}. However, the converse does not seem to hold true in general. In fact, in any domain ${\cal D} \subset S^3$ where $\sigma\,h^{-1}$ is not a symmetric matrix, equation \rf{sss1} gives $S_i^{(+),\, a}\,S_i^{(-),\, b} \neq 0 $. Therefore, in this domain we cannot have self-dual solutions, which satisfies \rf{self22}.  Additionally, note that this argument does not depend on whether $\tau$ is invertible.

In case of $G=SU(2)$ we can treat the Maurer-Cartan components $R_i^a$ as $3\times3$ matrix with the following ordering of rows and columns $R_{ia} \equiv R_i^a,\, i = 1,\, 2,\, 3$ and $a = 1, \,2, \, 3$. Therefore, $\varepsilon_{ijk} \,R_i^a\,R_j^b\,R_k^c=\varepsilon_{abc} \,\varepsilon_{ijk} \,R_{i1}\,R_{j2}\,R_{k3}= \varepsilon_{abc} \, \det R$, and using \rf{sigmadefmatrix}, we obtain
\be
\sigma_{ab} = -\frac{1}{2}\,\ve_{ijk}\,\varepsilon_{bcd}\,R_i^a\,R_j^c\,R_k^d = -\delta_{ab} \, \det R\ee
The $G=SU(2)$ case is very special since both $h^{-1}$ and $\sigma$ are symmetric by construction, and the structure constant reduces to the Levi-Civita symbol. Using these two properties, it was demonstrated in \cite{luiz1} that the self-duality equations \rf{self11} are a consequence of the static Euler-Lagrange equations associated with the $h$ fields. In particular, this implies that the static sector is equivalent to the self-dual sector for $G=SU(2)$.

\section{The holomorphic ansatz}
\setcounter{equation}{0}
\label{sec:holomorphic}

Let us consider a compact simple Lie group $G$, and let $\psi$ denote its highest positive root. This root can be write in terms of the the simple roots $\alpha_a$, $a=1,2,3,\ldots ,\, {\rm rank}\,G$, as $\psi=\sum_{a=1}^{{\rm rank}\,G} n_a\,\alpha_a$, where $n_a$'s are positive integers. The irreducible compact Hermitian symmetric spaces, as defined in (see \cite{helgason}), correspond to those cases where the expansion of $\psi$ in terms of  the simple roots presents at least one coefficient $n_a$ is equals to unity, i.e. 
\be
\psi = \alpha_{*}+\sum_{a=1\,,\, a\neq *}^{{\rm rank}\,G}n_a\,\alpha_a\;
\lab{alphastardef}
\ee
where $\alpha_{*}$ denote the simple root that appears only once in the expansion ($n_*=1$). 

Let us denote $\lambda_{*}$ the fundamental weight of $G$ which is not orthogonal to $\alpha_{*}$, i.e.
\be
\frac{2\,\lambda_{*}\cdot \alpha_{*}}{\alpha_{*}^2}=1\;;\qquad\qquad \qquad \frac{2\,\lambda_{*}\cdot \alpha_{a}}{\alpha_{a}^2}=0\;;\quad {\rm for}\quad a\neq * \lab{orthogonal}
\ee
The hermitian symmetric spaces are characterized by the $U(1)$ factor in the little group, and the involutive automorphism $\sigma$ ($\sigma^2=1$), defining the symmetric space structure is inner and constructed from the generator $\Lambda$ of the $U(1)$ subgroup, i.e.
\be
\sigma\(T\)\equiv e^{i\,\pi\,\Lambda}\, T\, e^{-i\,\pi\,\Lambda}\;;\qquad \qquad \Lambda\equiv\frac{2\,\lambda_{*}\cdot H}{\alpha_{*}^2}\;;\qquad \qquad \mbox{\rm for any}\;T\in {\cal G}
\lab{sigmadef}
\ee
where we choose to work in the Cartan-Weyl basis and $H_{i}$, $i=1,2,3,\ldots ,{\rm rank}\;G$, are the generators of the Cartan subalgebra of ${\cal G}$. Denoting $E_{\alpha}$ as the step operator associated to the root $\alpha$ of ${\cal G}$, the killing form of ${\cal G}$ become 
\be
{\rm Tr}\(H_i\,H_j\)=\delta_{ij}\,, \qquad\quad{\rm Tr}\(H_i\,E_{\alpha}\)=0\,,\qquad\quad{\rm Tr}\(E_{\alpha}\,E_{\beta}\)=\frac{2}{\alpha^2}\,\delta_{\alpha+\beta\,,\,0}
\lab{killingform}
\ee

The relations  \rf{sigmadef} and $\left[H_i,\,E_\alpha\right]=\alpha_i \,E_\alpha$, where the index $i$ denotes the component of the root $\alpha$, leads to $\left[\Lambda,\,E_\alpha\right] = \frac{2\,\lambda_{*}\cdot \alpha}{\alpha_*^2}\,E_{\alpha}$. Expanding the root through $\alpha= m_*\,\alpha_* + \sum_{a\neq *}^{{\rm rank}\,G} m_{\alpha_a}\,\alpha_a$, where $m_{\alpha_a}$ are integers and $m_*$ can take the values $-1,\,0,\,1$ due to \rf{alphastardef}, and using \rf{orthogonal} we obtain $\left[\Lambda,\,E_\alpha\right] = n_{\alpha_a*}\,E_{\alpha}$. Consequently, the step operators $E_{\pm\alpha}$ (anti)-commutes with $e^{i\,\pi\,\Lambda}$ for $n_{\alpha_a*}=0$ ($n_{\alpha_a*}=\pm 1$). Denoting $\gamma$ as any positive root of $G$ that does not contain $\alpha_{*}$ in its expansion in terms of simple roots, and $\alpha_{\kappa}$ as the remaining positive roots, we get from \rf{sigmadef} that
\be
\sigma\(H_i\)=H_i\;;\qquad\qquad \sigma\(E_{\pm \gamma}\)=E_{\pm \gamma}\;;\qquad\qquad \sigma\(E_{\pm \alpha_{\kappa}}\)=-E_{\pm \alpha_{\kappa}}
\lab{evenoddgenerators}
\ee
Therefore, the Lie algebra ${\cal G}$ of $G$ breaks in even and odd subalgebras under the involutive automorphism \rf{sigmadef}, denoted respectively by ${\cal K}$ and ${\cal P}$, i.e.  
\be
{\cal G}= {\cal P}+ {\cal K}\qquad\qquad {\rm with}\qquad \sigma\(P\) = -P\qquad \sigma\(K\)=K\qquad P\in {\cal P}\;;\quad K\in {\cal K}
\ee
Note that $\Lambda$ and $E_{\pm \gamma}$ belongs to the even subgroup ${\cal K}$, and $\Lambda$ generates an $U(1)_{\Lambda}$ invariant subalgebra of it. Consequently, we can write ${\cal K}={\cal H}\oplus {\Lambda}$, and we obtain the irreducible compact Hermitian symmetric space $G/H\otimes U(1)_{\Lambda}$. The subgroup $H$ is generated by $H_a\equiv \frac{2\,\alpha_a\cdot H}{\alpha_a^2}$, with $\alpha_a\neq \alpha_{*}$,  $\(E_{\gamma}+E_{-\gamma}\)$, and $i\,\(E_{\gamma}-E_{-\gamma}\)$. The odd subgroup is generated by $E_{\pm \alpha_{\kappa}}$, as defined in \rf{evenoddgenerators}, where $\kappa=1,\,2,...,\, \frac{{\rm dim}\;{\cal P}}{2}$.

The hermitian symmetric space have the form of a coset $G/K$, where $K$ is the little group $K=H\otimes U(1)$ and we get the usual algebraic structure of a symmetric space
\be
\sbr{{\cal G}}{{\cal G}}\subset {\cal G}\qquad\qquad\sbr{{\cal G}}{{\cal P}}\subset {\cal P}
\qquad\qquad\sbr{{\cal P}}{{\cal P}}\subset {\cal G}
\ee
The hermitian character of such symmetric spaces is that ${\cal P}$ is even dimensional and it is split by $\Lambda$ into two parts according the its eigenvalues
\be
{\cal P}={\cal P}_{+}+{\cal P}_{-}\qquad\qquad \sbr{\Lambda}{P_{\pm}}=\pm P_{\pm}\qquad\qquad P_{\pm}\in {\cal P}_{\pm}
\lab{hermitianniceprop}
\ee
The generators of ${\cal P}_{+}$ and ${\cal P}_{-}$ are respectively $E_{\alpha_{\kappa}}$ and $E_{-\alpha_{\kappa}}$, with $\kappa=1,2,\ldots ,\frac{{\rm dim}\,{\cal P}}{2}$. It turns out that ${\cal P}_{-}$ is like the hermitian conjugate of ${\cal P}_{+}$, and so both spaces have the same dimension, i.e. ${\rm dim}\,{\cal P_+}={\rm dim}\,{\cal P_-}=\frac{{\rm dim}\,{\cal P}}{2}$. Therefore, $\Lambda$ not only provides the automorphism $\sigma$, but it also provides a gradation of the Lie algebra ${\cal G}$ into subspaces of grades $0$ and $\pm 1$. Since there are no subspaces of grades $\pm 2$, it turns out that ${\cal P}_{\pm}$ are abelian. So we have
\be
 \sbr{{\cal K}}{{\cal K}}\subset {\cal K}\qquad  \sbr{{\cal K}}{{\cal P}_{\pm}}\subset {\cal P}_{\pm}\qquad 
 \sbr{{\cal P}_{+}}{{\cal P}_{+}}=\sbr{{\cal P}_{-}}{{\cal P}_{-}}=0\qquad \sbr{{\cal P}_{+}}{{\cal P}_{-}}\subset {\cal K}
 \lab{finerstructure}
 \ee
The compact irreducible hermitian symmetric spaces are
\be
\begin{array}{@{\extracolsep{30pt}}ll}
SU(p+q)/SU(p)\otimes SU(q)\otimes U(1) \,;& SO(2N)/SU(N)\otimes U(1)\,; \\ 
SO(N+2)/SO(N)\otimes U(1) \,; & Sp(N)/SU(N)\otimes U(1)\,;\\
E_6/SO(10)\otimes U(1)\,;& E_7/E_6\otimes U(1)
\end{array}
\ee
The trace form is invariant under the automorphism $\sigma$, i.e.
 ${\rm Tr}\(\sigma\(T\)\,\sigma\(T^{\prime}\)\)={\rm Tr}\(T\,T^{\prime}\)$. Therefore, the even and odd generators are orthogonal
 \be
 {\rm Tr}\({\cal P}\,{\cal K}\)=0 \lab{ortho1}
 \ee
 In addition one has
 \be
 0={\rm Tr}\(\Lambda\,\sbr{{\cal P}_{\pm}}{{\cal P}_{\pm}}\)={\rm Tr}\({\cal P}_{\pm}\,\sbr{\Lambda}{{\cal P}_{\pm}}\)=\pm{\rm Tr}\({\cal P}_{\pm}\,{\cal P}_{\pm}\) \lab{generatorpm}
 \ee
 and so
 \be
 {\rm Tr}\({\cal P}_{+}\,{\cal P}_{+}\)={\rm Tr}\({\cal P}_{-}\,{\cal P}_{-}\)=0
 \ee
 The even subalgebra ${\cal K}$ has the form ${\cal K}={\cal H}\oplus \Lambda$. If ${\cal H}$ is simple or even semi simple (no $U(1)$ factors), then it is true that any of its elements can be written as the commutator of some other two, i.e. ${\cal H}=\sbr{{\cal H}^{\prime}}{{\cal H}^{\prime\prime}}$. Then it follows that
 \be
 {\rm Tr}\(\Lambda\, {\cal H}\)={\rm Tr}\(\Lambda\,\sbr{{\cal H}^{\prime}}{{\cal H}^{\prime\prime}}\)=
 {\rm Tr}\(\sbr{\Lambda}{{\cal H}^{\prime}}\,{\cal H}^{\prime\prime}\)=0
 \ee
 and so
 \be
 {\rm Tr}\(\Lambda\, {\cal H}\)=0
 \lab{tracelambdah}
 \ee

One nice thing about symmetric spaces (not only hermitian) is that one can parameterize it quit easily. Given a matrix of the group $G$ one construct the so-called principal variable
\be
X\(g\)=g\,\sigma\(g\)^{-1}\qquad \mbox{\rm and so}\qquad X\(g\, k\)=X\(g\)\qquad {\rm and}\qquad \sigma\(X\)=X^{-1}
\lab{xdef}
\ee
with $k$ being any element of the $K$ subgroup and $g$ is any element of $G$. Therefore, $X\(g\)$ parameterize the coset space (symmetric space) $G/K$. The three dimensional space $\IR^3$ can be foliated with spheres with center at the origin, being useful introduce the spherical coordinates $\(r\,,\,\theta\,,\, \vp\)$. We stereo graphically project the spheres on a plane with the infinity identified to a point, i.e, the Riemann sphere. The maps from that sphere to the hermitian symmetric space are labeled by integers. Let $z$ and ${\bar z}$ be the complex coordinates on that plane introduced by the coordinate system $\(r\,,\,z\,,\,{\bar z}\)$ defined by ($z=z_1+iz_2$)
\be
x_1= r\, \frac{i\(\bar{z}-z\)}{1+\mid z\mid^2} \; ; \qquad 
x_2= r\, \frac{z+\bar{z}}{1+\mid z\mid^2} \; ; \qquad 
x_3= r\, \frac{\(-1+\mid z\mid^2\)}{1+\mid z\mid^2} \lab{coordinates}
\ee
which have the metric
\be
ds^2=dr^2+ \frac{4\,r^2}{\(1+\mid z\mid^2\)^2} \, dz\,d{\bar z} \lab{metric}
\ee

A powerful ansatz for the Skyrme field $U$ was proposed by L. A. Ferreira and L. R. Livramento in  \cite{Ferreira:2024ivq}. First, it consider the rational map for the group elements $U$ of the group $G$ as 
\be
U= g\, e^{i\,f\(r\)\, \Lambda}\, g^{-1}=e^{i\,f\(r\)\,g\,\Lambda\,g^{-1}}\qquad\qquad \qquad \sigma\(g\)=g^{-1} \lab{holog}
\ee
where $f\(r\)$ is a radial profile function, $\Lambda$ is defined in \rf{sigmadef} and $g$ is  an element of the compact Lie group $G$, and it depends only $z$ and ${\bar z}$, i.e. $g=g\(z\,,\,{\bar z}\)$. Clearly, as $\Lambda$ commutes with the elements of the even subgroup $K=H\otimes U(1)_{\Lambda}$, it so follows that $g\, k\,\Lambda\,\,\(g\,k\)^{-1}= g\,\Lambda\,g^{-1}$, with $k\in K$. Consequently, the term $g\,\Lambda\,g^{-1}$ depends only on the fields parameterizing the cosets in $G/H\otimes U(1)_{\Lambda}$, which are parametrized by a principal variable of the form \rf{xdef}. In fact, we can take $g$ as a principal variable \rf{xdef}, i.e. $g\(w\)=w\,\sigma\(w\)^{-1}$, with $w$ being any element of $G$ depending only on $z$ and ${\bar z}$. The definition \rf{xdef} implies that $\sigma\(g\)=g^{-1}$, and so the principal variable $X$ is reduced to $X\(g\)=g\,\sigma\(g\)^{-1}=g^2$.

The Maurer-Cartan form associated to $g$, i.e. $g^{-1}\,\pa_i g$ can be projected into even and odd subspaces through
\be
g^{-1}\,\pa_i g=P_i+K_i\qquad\qquad
P_i= \frac{1-\sigma}{2} g^{-1}\,\pa_i g\qquad\qquad
K_i= \frac{1+\sigma}{2} g^{-1}\,\pa_i g \lab{ws}
\ee
Since we are dealing with hermitian symmetric spaces we can use \rf{hermitianniceprop} to split $P_i$ into the $\pm 1$ subspaces, i.e. 
\be
P_i=P^{(+)}_i+P^{(-)}_i\qquad\qquad \sbr{\Lambda}{P^{(\pm)}_i}=\pm P^{(\pm)}_i\lab{wpm}
\ee

The second key ingredient of the ansatz proposed in \cite{Ferreira:2024ivq} is the construction of the $g$ elements. Let us introduce 
\be
S=\sum_{\kappa} w_{\kappa}\,E_{\alpha_{\kappa}}\;; \qquad\qquad \sbr{\Lambda}{S}=S
\lab{sdef}
\ee
which lies on ${\cal P}_+$ by definition, and where $w_{\kappa}$ are functionals of the fields parameterizing the hermitian symmetric spaces $G/H\otimes U(1)_{\Lambda}$. Consequently, 
\be
S^{\dagger}=\sum_{\kappa} w_{\kappa}^*\,E_{-\alpha_{\kappa}}\;; \qquad\qquad \sbr{\Lambda}{S^{\dagger}}=-S^{\dagger}
\lab{sdaggerdef}
\ee

First, $S$ and $S^\dagger$ are even respectively holormorphic and anti-holomorphic, i.e. $S=S(z)$ and $S^\dagger(\bar{z})$, or vice versa. We can consider both cases by writing
\be S=S(\chi) \; \qquad {\rm and} \qquad S^\dagger =S^\dagger(\bar{\chi})\qquad\qquad {\rm with}\qquad\qquad\chi=z,\,\bar{z} \lab{holomorphics}\ee 
where $\chi=z$ ($\chi=\bar{z}$) corresponds to a (anti-)holomorphic matrix $S$. Second, the ansatz works for representations of the Lie algebra of $G$ where the matrix associated to $S$ is nilpotent with index of nilpotency equal to two, and an eigenvector of the hermitian matrix $S\,S^{\dagger}$ with a non-negative eigenvalue, i.e.
\be
S^2=0\;;\qquad\qquad \qquad \qquad\(S\,S^{\dagger}\)\,S=\omega\,S
\lab{omegadef}
\ee
where the eigenvalue $\omega$ is non-negative. From \rf{omegadef} we can obtain some useful relations, such as 
\be \pa_i S\,S=\pa_i S^\dagger\,S^\dagger=0;\qquad\qquad S\,S^\dagger\,\pa_i S\,S^\dagger\,S =\omega\,\(\pa_i \omega \,S-S\,\pa_iS^\dagger\, S\) \lab{usefulrel1}\ee
On the other hand, the equation \rf{omegadef} implies ${S^{\dagger}}^2=0$ and $\(S^{\dagger}\,S\)\,S^{\dagger}=\omega\,S^{\dagger}$. Third, the ansatz for $g=g(z,\,\bar{z})$ corresponds to 
\be
g =   \one +\frac{1}{\vartheta}\, \left[i\(S+S^\dagger\) -\frac{1}{\vartheta+1} \,\left(S\,S^\dagger+S^\dagger\,S\right) \right] ; \qquad  \qquad\vartheta\equiv \sqrt{1+\omega}
 \lab{gdef2}
\ee
which is unitary and satisfies $\sigma\(g\)=g^{-1}$. 
Alternatively, $g$ given by \rf{gdef2} can be also written as
\be g=e^{i\,S}\,e^{\varphi\,\sbr{S}{S^{\dagger}}}\,e^{i\,S^{\dagger}}=e^{i\,a\,\(S+S^\dagger\)}  \lab{galternative}\ee
with $\varphi=\omega^{-1}\,\ln\sqrt{1+\omega}$ and $a= \omega^{-\frac{1}{2}}\,{\rm arcsin}\(\frac{\sqrt{\omega}}{\sqrt{1+\omega}}\)$. Using also \rf{ws}, \rf{wpm} and \rf{omegadef} we get  
\br
 P_i^{(+)} &=& \frac{i}{\vartheta}\,\pa_iS+\frac{i}{\vartheta^2\,\(1+\vartheta\)}\(S\,\pa_i S^{\dagger}\, S-2\,\vartheta\,\pa_i \vartheta\,S \) \nonumber\\
P_i^{(-)} &=& \frac{i}{\vartheta}\,\pa_iS^{\dagger} +\frac{i}{\vartheta^2\,\(1+\vartheta\)}\(S^{\dagger}\,\pa_i S\, S^{\dagger}- 2\,\vartheta\,\pa_i \vartheta\,S^\dagger\)  \lab{pi}
\er
which satisfies $(P_i^{(+)})^\dagger = - P_i^{(-)}$. Let us introduce the hermitian and invertible operator
\be \Omega \equiv \one + \frac{S\,S^\dagger+ S^\dagger\,S}{1+\vartheta} \qquad \Rightarrow \qquad \Omega^{-1} =\one - \frac{S\,S^\dagger+S^\dagger\,S}{\vartheta\,\(1+\vartheta\)} \lab{omegas}\ee
Using also \rf{omegadef} and \rf{usefulrel1}, it so follows that 
\be \Omega\,P_i^{(+)}\,\Omega = i\,\pa_i S \;\qquad\qquad \stackrel{\dagger}{\Rightarrow}\qquad\qquad \Omega\,P_i^{(-)}\,\Omega = i\,\pa_i S^\dagger \lab{extraction}\ee
The details of these calculations are presented in \cite{Ferreira:2024ivq}.

An important consequence of \rf{extraction} is that the holomorphic and the holomorphic ansatz \rf{holomorphics} are equivalent to
\be
\pa_{\bar{\chi}} S=\pa_{{\chi}} S^\dagger=0\quad \Leftrightarrow\quad  P_{\bar{\chi}}^{(+)}=P_{{\chi}}^{(-)}=0 \quad\qquad  \mbox{with} \quad\qquad \chi\equiv \left\{\begin{array}{ll} z, & S=S(z)\\ \bar{z}, & S=S(\bar{z})\end{array} \right. \lab{wg}
\ee
In addition, for later convenience we can also introduce the sign function 
\be \eta \equiv \left\{\begin{array}{ll} +1, & \chi = z\\ -1, & \chi=\bar{z}\end{array} \lab{etadef}\right.\ee
The relation \rf{wg}  also shows that the terms $S\,\pa_\chi S^\dagger \, S$ and $S^\dagger\,\pa_{\bar{\chi}} S\, S^\dagger$ vanishes, reducing $P_{{\chi}}^{(+)}$ and $P_{\bar{\chi}}^{(-)}$, which can be obtained from \rf{pi}, to
\be  P_\chi^{(+)} =i\,\frac{\(1+\vartheta\)^2}{\vartheta}\,\pa_\chi \(\frac{S}{\(1+\vartheta\)^2}\);\qquad\qquad P_{\bar{\chi}}^{(-)}= i\,\frac{\(1+\vartheta\)^2}{\vartheta}\,\pa_{\bar{\chi}} \(\frac{S^\dagger}{\(1+\vartheta\)^2}\)\lab{pir}\ee

Using \rf{sigmadef} and \rf{gdef2} we have that the Lie algebra element appearing in the ansatz \rf{holog} is given by 
\be
g\,\Lambda\,g^{-1}=\Lambda -\frac{1}{\(1+\omega\)}\left(\sbr{S}{S^{\dagger}}+i\(S-S^{\dagger}\)\right)
\lab{bitofanasatz}
\ee
In particular, if the square of the $U(1)$ generator $\Lambda$ can be written as $\Lambda^2=c\,\Lambda + \frac{1}{4}\,\(1-c^2\)\,\one$ in such a representation, where $c$ is a real number, it so follows that the quantity 
\be Z \equiv \frac{1+c}{2}\,\one -g\,\Lambda\,g^{-1} \lab{projector0}\ee
is a projector, i.e. $Z=Z^2$. This allow us to easily compute the exponential on the r.h.s. of \rf{holog}reducing the rational map to
\be 
U= e^{if\,\frac{1+c}{2}\,\one}\,e^{-i\,f\,Z}= e^{if\,\frac{c+1}{2}}\,\left[\one+\(e^{-if}-1\)Z\right]
\lab{Uexpanded}\ee
Although this condition on $\Lambda$ is not assumed in the above ansatz, it appears in the parameterization of the hermitian symmetric space $G/K=SU\(p+q\)/SU\(p\)\otimes SU\(q\)\otimes U\(1\)$ given in \cite{Ferreira:2024ivq}.

\subsection{The self-dual sector}

Using \rf{ws}, \rf{wpm} and the fact that  $K_i$ commutes with $\Lambda$, the Maurer-Cartan form $R_i =i\,\pa_i U\,U^{-1}$ associated to the rational map \rf{holog} becomes
 \be
R_{i} = -V^{-1}\, \Sigma_i\, V \lab{rv}
\ee
with 
\br
V\equiv e^{-i\,f\,\Lambda/2}\,g^{-1} ;\quad\qquad \qquad\Sigma_i&\equiv &\partial_if\,\Lambda
- 2\,\sin \frac{f}{2}\(P^{(+)}_i-P^{(-)}_i\)
\lab{Sigmadef}
\er
We now have that
\be
R_i^b= \frac{1}{\kappa}\,{\rm Tr}\(T_b \, R_i\)=-\frac{1}{\kappa}\,{\rm Tr}\(V\, T_b\,V^{-1}\, \Sigma_i\)=
-\frac{1}{\kappa}\,{\rm Tr}\( T_c\, \Sigma_i\)\,d_{cb}\(V\) \lab{Rsigma}
\ee
where we have introduced the matrix for the group elements in the adjoint representation of $G$
\be
g\, T_a\, g^{-1}=T_{b}\,d_{ba}\(g\)
\ee
Similarly we have
\be
{\rm Tr}\(T_a\sbr{R_j}{R_k}\)= {\rm Tr}\(V\,T_a\, V^{-1}\,\sbr{\Sigma_j}{\Sigma_k}\)=
 {\rm Tr}\(T_c\, \sbr{\Sigma_j}{\Sigma_k}\)\, d_{ca}\(V\)
\ee
From \rf{Sigmadef} we note that $\Sigma_i$ has components along the $U(1)$ generator $\Lambda$, and on the odd subspace ${\cal P}$. In fact, using \rf{wg} and the definition \rf{Sigmadef} we have
\br
\Sigma_r = f' \,\,\Lambda \,;\quad \qquad \Sigma_\chi = - 2\,\sin \frac{f}{2}\,P^{(+)}_\chi\,;\quad\qquad \Sigma_{\bar{\chi}} =  2\,\sin \frac{f}{2}\,P^{(-)}_{\bar{\chi}} \lab{sigmasr}
\er
where $P^{(+)}_\chi$ and $P^{(-)}_{\bar{\chi}}$ are given by \rf{pir}.

It follows from \rf{wpm} and \rf{rv} that  $ \ve_{ijk}\,{\rm Tr}\(R_i\,R_j\,R_k\) =12\,\ve_{ijk}\,\partial_if\,\sin^2\frac{f}{2}\,{\rm Tr}\(P^{(+)}_j\,P^{(-)}_k\)$, which reduces the topological charge \rf{topcharge} to 
\br
 Q&=& \frac{i}{4\,\pi^2\,\kappa}\int dr\,dz\,d{\bar z}\;\partial_rf\,\sin^2\frac{f}{2}\,{\rm Tr}\(P^{(+)}_z\,P^{(-)}_{{\bar z}}-P^{(+)}_{{\bar z}}\,P^{(-)}_z\) 
 \nonumber\\ 
 &=&\frac{1}{2\,\pi}\,\left[f\(r\)-\sin f\(r\)\right]_{r=0}^{r=\infty}\;Q_{{\rm top}}
 \lab{pretopcharge3d}
\er
with $Q_{{\rm top}}\equiv \frac{i}{4\,\pi\,\kappa}\,
 \int dz\,d\bar{z}\;{\rm Tr}\(P^{(+)}_z\,P^{(-)}_{{\bar z}}-P^{(+)}_{{\bar z}}\,P^{(-)}_z\)=\eta\,\frac{i}{4\,\pi\,\kappa}\, \int dz\,d\bar{z}\;{\rm Tr}\(P^{(+)}_\chi\,P^{(-)}_{{\bar \chi}}\)$, where we use \rf{wg} and \rf{etadef}.

 Using \rf{rv} and \rf{sigmasr} the self-duality equations \rf{selfdualityeqs2} can be written as
\be
\lambda\, \widetilde{\tau}_{cb}\,\tilde{h}_{ba}=\widetilde{\sigma}_{ca}
\lab{selfdualityeqs3}
\ee
where we have introduced the matrices
\be
{\tilde h}_{ab}\equiv d_{ac}\(V\)\, h_{cd}\, d^{-1}_{db}\(V\) 
 \; ;\quad {\widetilde \tau}_{ab}\equiv d_{ac}\(V\)\, \tau_{cd}\, d^{-1}_{db}\(V\) \; ;\quad {\widetilde \sigma}_{ab}\equiv d_{ac}\(V\)\, \sigma_{cd}\, d^{-1}_{db}\(V\) 
\lab{htildedef}
\ee
The adjoint representation of a compact simple Lie group is unitary and real, and so $d$ is an orthogonal matrix, i.e. $d^T=d^{-1}$. Therefore, ${\tilde h}_{ab}$ and ${\widetilde \tau}_{ab}$ are still symmetric. In addition, we have $\widetilde{\tau}_{ab} = \Sigma_i^a\,\Sigma_i^b$ and $\widetilde{\sigma}_{a b} = -\frac{i}{2}\,\trace\(T_a\,\Sigma_i\)\,\ve_{ijk}\,\trace\(T_b\sbr{\Sigma_j}{\Sigma_k}\)$. Using $i\,\ve_{ijk}\,\frac{\pa r}{\pa x^i}\,\frac{\pa \chi}{\pa x^j}\,\frac{\pa \bar{\chi}}{\pa x^k} = \eta\,\frac{\(1+\mid z \mid^2\)^2}{2\,r^2}$ and \rf{sigmasr} we obtain
\br
\widetilde{\sigma}_{a b}& \equiv & y\,\(\Gamma_{r \chi \bar{\chi}}^{ab} + \Gamma_{\bar{\chi} r \chi }^{ab} +\Gamma_{\chi \bar{\chi} r}^{ab} \)\nonumber
\er
with $y \equiv  2\,\eta\,f'\,\sin^2 \frac{f}{2}\,\frac{\(1+\mid z \mid^2\)^2}{r^2}$ and  $\Gamma_{\alpha\beta\gamma}^{ab}\equiv -\(4\,f'\,\sin^2 \frac{f}{2}\)^{-1}\,\trace\(T_a\,\Sigma_\alpha\)\,\trace\(T_b\sbr{\Sigma_\beta}{\Sigma_\gamma}\)$, which is antisymmetric under the exchange of its last two indices $\beta$ and $\gamma$ and has the following components
\br
{\Gamma}_{r \chi \bar{\chi}}^{ab} & = &  \trace\(T_a\,\Lambda\)\,\trace\(T_b\, \left[P^{(+)}_{{\chi}}  ,\, P^{(-)}_{\bar{\chi}} \right]\) \, \:\:\:\, \qquad\nonumber {\rm that \:\: vanishes \:\: if \:\: } T_a\neq \Lambda {\rm \:\: or \:\:}T_b \in {\cal P}\\
{\Gamma}_{\bar{\chi} r \chi }^{ab} &=& \trace\(T_a\,P^{(-)}_{\bar{\chi}}\)\,\trace\(T_b\, P^{(+)}_\chi\) \, \qquad\;\quad\qquad \nonumber {\rm that \:\: vanishes \:\: if \:\: } T_a\notin {\cal P_+} {\rm \:\: or \:\:}T_b \notin {\cal P_-}  \\
{\Gamma}_{\chi\bar{\chi} r  }^{ab} &=& \trace\(T_a\, P^{(+)}_\chi\)\,\trace\(T_b\, P^{(-)}_{\bar{\chi}}\) \, \qquad\;\quad\qquad \nonumber {\rm that \:\: vanishes \:\: if \:\: } T_a\notin {\cal P_-} {\rm \:\: or \:\:}T_b \notin {\cal P_+}  
 \er
where we use \rf{wpm} and the cyclic property of the trace. The components of the matrices $\widetilde{\tau}$ and $\widetilde{\sigma}$ become
\br
\widetilde{\tau}_{{\cal H} \, a} & =  & 0\;;\;\;\qquad \qquad \qquad \widetilde{\tau}_{\Lambda a} = \left\{\begin{array}{ll} f^{\prime 2}\,\left[\trace\(\Lambda^2\)\right]^2\, , & T_a = \Lambda\\ 0, & {\rm otherwise}\end{array} \right. \nonumber\\
\widetilde{\tau}_{{\cal P}_+ \, a}& = & \left\{\begin{array}{ll}  \(-\eta\,f'\)^{-1}\, y\,\trace\({\cal P}_+\,P^{(-)}_{\bar{\chi}}\)\,\trace\(T_a\,P^{(+)}_\chi\)\, , & T_a \in  {\cal P}_-\\ 0, & {\rm otherwise}\end{array} \right. \lab{taus}\\
\widetilde{\tau}_{{\cal P}_- \, a}& = & \left\{\begin{array}{ll}  \(-\eta\,f'\)^{-1}\,y\,\trace\({\cal P}_-\,P^{(+)}_\chi\)\,\trace\(T_a\,P^{(-)}_{\bar{\chi}}\)\, , & T_a \in  {\cal P}_+\\ 0, & {\rm otherwise}\end{array} \right. \nonumber
\er
and 
\br
\widetilde{\sigma}_{a\,{\cal H}} &= &\left\{\begin{array}{ll} y\,\trace\(\Lambda^2\)\,\trace\({\cal H}\, \left[P^{(+)}_{{\chi}}  ,\, P^{(-)}_{\bar{\chi}} \right]\)\, , & T_a = \Lambda\\ 0, & {\rm otherwise}\end{array} \right. \,\nonumber\\
\widetilde{\sigma}_{\Lambda\Lambda} &=& y\,\trace\(\Lambda^2\)\,\trace\(P^{(+)}_\chi\,P^{(-)}_{\bar{\chi}}\)\, \nonumber \\
\widetilde{\sigma}_{{\cal P}_+{\cal P}_-} &=& \widetilde{\sigma}_{{\cal P}_-{\cal P}_+} = y\,\,\trace\({\cal P}_+\, P^{(-)}_{\bar{\chi}}\)\,\trace\({\cal P}_- P^{(+)}_\chi\) \,\ \stackrel{\rf{taus}}{=} -\eta\,f'\, \widetilde{\tau}_{{\cal P}_+ \, {\cal P}_-}
\lab{preeq} \\
\widetilde{\sigma}_{{\cal H}\,b} & = & \widetilde{\sigma}_{{\cal P} \Lambda }=\widetilde{\sigma}_{{\cal P}_+{\cal P}_+}=\widetilde{\sigma}_{{\cal P}_-{\cal P}_-}=0 \nonumber
\er
The index ${\cal H}$ in the row or column indices of the matrices $\widetilde{\tau}$ and $\widetilde{\sigma}$ represents any index $a$  that labels the generators $T_a$ of the subalgebra ${\cal H}$, and so on. A crucial consequence of  ${\tilde \tau}_{a \, {\cal H}}={\widetilde \sigma}_{a\,{\cal H}}=0$ for all $a=1,\,...,\,\dim G$ is that none of the self-dual equations \rf{selfdualityeqs3} depends on the  $\tilde{h}_{{\cal H}{\cal H}}$ fields. Then, if ${\cal H} \neq \emptyset$, it follows that ${\widetilde h}_{{\cal H}{\cal H}}$ is totally undetermined and ${\widetilde \tau}$ is not invertible. Otherwise, we could apply $\widetilde{\tau}^{-1}$ to the self-dual equations \rf{selfdualityeqs3} fixing $\tilde{h}$ interely.

The self-duality equations  associated with the row index ${\cal H}$ of $\widetilde{\tau}$, given by $\lambda\, \widetilde{\tau}_{{\cal H}\,b}\,\tilde{h}_{b\,a}=\widetilde{\sigma}_{{\cal H}\,a} $, are automatically satisfied by \rf{taus} and \rf{sigmasr}, and the remaining equations are reduced to
\br
\widetilde{\sigma}_{\Lambda \, b}& = & \lambda\, \widetilde{\tau}_{\Lambda\Lambda}\,\tilde{h}_{\Lambda\, b} \quad \qquad \Rightarrow \quad \qquad  \tilde{h}_{\Lambda\,b} = \frac{\widetilde{\sigma}_{\Lambda \, b}}{\lambda\, \widetilde{\tau}_{\Lambda\Lambda}}\qquad \lab{2} \\
0& = & \widetilde{\tau}_{{\cal P}_+\,{\cal P}_-}\,\tilde{h}_{{\cal P}_-\, {\cal P}_+}=\widetilde{\tau}_{{\cal P}_+\,{\cal P}_-}\,\tilde{h}_{{\cal P}_-\, {\cal H}}=\widetilde{\tau}_{{\cal P}_-\,{\cal P}_+}\,\tilde{h}_{{\cal P}_+\, {\cal P}_-}=\widetilde{\tau}_{{\cal P}_-\,{\cal P}_+}\,\tilde{h}_{{\cal P}_+\, {\cal H}}  \lab{3}  \\
0 & = & \widetilde{\tau}_{{\cal P}_-\,{\cal P}_+}\, \(\tilde{h}_{{\cal P}_+\, {\cal P}_+}+\eta\,\lambda^{-1}\,f'\,\one\) = \widetilde{\tau}_{{\cal P}_+\,{\cal P}_-}\, \(\tilde{h}_{{\cal P}_-\, {\cal P}_-}+\eta\,\lambda^{-1}\,f'\,\one\) \lab{4}  
\er
Note that there is an implicit sum over the line index of the $\tilde{h}$ matrix, leading to a linear system to the $\tilde{h}_{ab}$ fields.  However, in \rf{2} this sum is performed over a single generator, which corresponds to the $U(1)$ generator $\Lambda$. This is an consequence of the fact that $\widetilde{\tau}_{\Lambda\,\Lambda}$ is the only non-vanishing component of $\widetilde{\tau}_{\Lambda\, a}$ given in \rf{taus}. Therefore, $\widetilde{h}_{\Lambda\,a}$ is fully determined by
\be {\tilde h}_{\Lambda \,\Lambda}\:\,=\alpha\,\eta\,{\rm Tr}\(P^{(+)}_\chi\,P^{(-)}_{\bar{\chi}}\);
\lab{hl} \quad {\tilde h}_{\Lambda \,{\cal H}} = \alpha\,\eta\,{\rm Tr}\({\cal H} \,\sbr{P^{(+)}_\chi}{P^{(-)}_{\bar{\chi}}}\)
;\quad {\tilde h}_{\Lambda \,{\cal P}_{\pm}}=0 \ee
\be {\rm with}\qquad\qquad \alpha \equiv \frac{2\,\sin^2\frac{f}{2}}{\lambda\,f'\,{\rm Tr}\( \Lambda^2\)}\,\frac{\(1+\mid z\mid^2\)^2}{r^2}\nonumber \ee
Note that we can raplace the modified trace $\trace$ defined in \rf{normalizedtr} by the usual trace $\Tr$ since the $\kappa$ factor cancels in the self-duality equations \rf{selfdualityeqs3}.

Clearly, if the matrix $\widetilde{\tau}_{{\cal P}_-\,{\cal P}_+}$ is invertible, then \rf{3} and \rf{4} leads to 
\br
\tilde{h}_{{\cal P}_\pm\, {\cal H}} & = & 0\;\qquad\qquad \tilde{h}_{{\cal P}\, {\cal P}}  = - \eta\,\lambda^{-1}\,f'\,\one \lab{5}
\er
However, if $\widetilde{\tau}_{{\cal P}_-\,{\cal P}_+}$ is not invertible, the fields \rf{hl} and \rf{5} are still a particular self-dual solution of \rf{selfdualityeqs3}. In any case, using \rf{taus}, the self-duality equation for the fields $\tilde{h}_{{\cal P}_\pm\, {\cal H}}$ of  \rf{3} can be written as 
\be {\rm Tr}\({\cal P}_{-}\,P^{(+)}_\chi\)\,{\tilde h}_{{\cal P}_{-}\,{\cal H}}={\rm Tr}\( {\cal P}_{+}\,P^{(-)}_{\bar{\chi}}\)\,{\tilde h}_{{\cal P}_{+}\,{\cal H}}=0 \lab{ph} \ee 
There is no other self-duality equation that depends on the fields ${\tilde h}_{{\cal P}_{\pm}\,{\cal H}}$. Thus, there are just $\dim {\cal H}$ equations to fix the components ${\tilde h}_{{\cal P}_{+}\,{\cal H}}$, and there is an independent set of $\dim {\cal H}$ equations to fix the components ${\tilde h}_{{\cal P}_{-}\,{\cal H}}$. Therefore, there are at least $2\,{\rm dim}\,{\cal H}\,({\rm dim}\,{\cal P}_+-1)$ components of $\tilde{h}_{{\cal H}{\cal P}}$ free. On the other hand, using also \rf{wg}, we obtain a set of $\dim {\cal P}_+$ equations to the fields ${\tilde h}_{{\cal P}_{-}\,{\cal P}_+}$ given by
\be {\rm Tr}\({\cal P}_{-}\,P^{(+)}_\chi\)\,{\tilde h}_{{\cal P}_{-}\,{\cal P}_+}=0  \lab{ppm} 
\ee
There is another set of linear equations given by ${\rm Tr}\( {\cal P}_{+}\,P^{(-)}_{\bar{\chi}}\)\,{\tilde h}_{{\cal P}_{+}\,{\cal P}_-}=0$ that comes from \rf{3} but corresponds to the complex conjugate of \rf{ppm}. Finally, using \rf{taus} we can write \rf{4} as
\br
{\rm Tr}\( {\cal P}_{+}\,P^{(-)}_{\bar{\chi}}\)\,{\tilde h}_{{\cal P}_{+}\,{\cal P}_{+}}&=& -\eta\, \lambda^{-1}\,f'\,
{\rm Tr}\({\cal P}_{+} \,P^{(-)}_{\bar{\chi}}\)\lab{ppf}\\
{\rm Tr}\( {\cal P}_{-}\,P^{(+)}_\chi\)\,{\tilde h}_{{\cal P}_{-}\,{\cal P}_{-}} &=& -\eta\, \lambda^{-1}\,f'\,
{\rm Tr}\({\cal P}_{-} \,P^{(+)}_{{\chi}}\)\lab{mmf}
\er  
Consequently, there are $\dim {\cal P}_+$ equations to fix the $\dim {\cal P}_+\,(\dim {\cal P}_+ +1)/2$ fields ${\tilde h}_{{\cal P}_{+}\,{\cal P}_{+}}$, and the same follows for the fields ${\tilde h}_{{\cal P}_{-}\,{\cal P}_{-}}$. Since ${\tilde h}_{{\cal P}{\cal P}}$ has ${\rm dim}\,{\cal P}$ diagonal elements and forms itself a symmetric matrix, then the relations \rf{ppm}, \rf{ppf} and \rf{mmf} together compose a set of $3\,{\rm dim}\,{\cal P}_+$ equations that contains $\frac{{\rm dim}\,{\cal P}\,\({\rm dim}\,{\cal P}+1\)}{2} = {\rm dim}\,{\cal P}_+\,\(2\,{\rm dim}\,{\cal P}_+ + 1\)$ components of the $\tih$-fields. Such facts lead to the freedom of at least $2\,{\rm dim}\,{\cal P}_+\,\({\rm dim}\,{\cal P}_+-1\)$ components of the $\tilde{h}_{{\cal P}{\cal P}}$ matrix. 

The above arguments show that ${\rm dim}\,{\cal P}_+ = 1$ is a necessary condition for the fields $\tilde{h}_{{\cal P}{\cal P}}$ to be fully determined by the self-duality equations \rf{selfdualityeqs3}. On the other hand, we also show above that if $\widetilde{\tau}_{{\cal P}_-\,{\cal P}_+}$ is invertible, then $\tilde{h}_{{\cal P}{\cal P}}$ must be fully determined by \rf{5}. Consequently, ${\rm dim}\,{\cal P}_+ = 1$ is also a necessary condition for the $\widetilde{\tau}_{{\cal P}_-\,{\cal P}_+}$ matrix be invertible. In particular, for the $G=SU(2)$ case $\widetilde{\tau}_{{\cal P}_-\,{\cal P}_+}$ is a real-value function.

\section{The $SU(2)/U(1)$ Hermitian symmetric space}
\label{sec:su2}
\setcounter{equation}{0}

In this case we have the symmetric space $SU(2)/U(1)$ and so 
\be
\Lambda = T_3\,;\qquad\qquad {\cal H}=\emptyset\,;\qquad\qquad {\cal P}_{+}=\{T_{+}\}\,;\qquad\qquad {\cal P}_{-}=\{T_{-}\} \lab{su2generators}
\ee
with 
\be
\sbr{T_3}{T_{\pm}}=\pm T_{\pm}\,;\qquad\qquad \qquad \sbr{T_{+}}{T_{-}}=2\,T_3
\ee
The quantity $g$ is
\br
g=\frac{1}{\sqrt{1+\mid u\mid^2}}\,\(\begin{array}{cc}
1&i\,u\\
i\,{\bar u}&1
\end{array}\)
\er
with $S=u$ and $S^\dagger=\bar{u}$ and 
\br
g^{-1}\,\partial_i g=\frac{1}{1+\mid u\mid^2}\left[i\(\partial_iu\,T_{+}+\partial_i{\bar u}\,T_{-}\) +\(u\,\partial_i{\bar u} - {\bar u}\,\partial_i u\)\,T_3\right]
\er
Then, the quantities $K_i$, $P_i^{(+)}$ and $P_i^{(-)}$ introduced in \rf{ws} and \rf{wpm} become 
\be
K_i=\frac{u\,\partial_i{\bar u}-{\bar u}\,\partial_i u}{1+\mid u\mid^2}\,T_3\,; \qquad\qquad P^{(+)}_i= \frac{i\,\partial_iu}{1+\mid u\mid^2}\,T_{+} \,;\qquad\qquad 
P^{(-)}_i= \frac{i\,\partial_i{\bar u}}{1+\mid u\mid^2}\,T_{-} \lab{wsu2}
\ee
Note that (anti)-holomorphic ansatz $S=S\(\chi\)$ implies $u = u\(\chi\)$. We  shall use the trace form in the doublet representation where (see \rf{kappadef})
\be
\kappa=\frac{1}{2}\,;\qquad\qquad\qquad {\rm Tr}\,\(T_{+}\,T_{-}\)=1
\ee

Clearly, since there is no generator of the subalgebra ${\cal H}$, the self-dual equations \rf{ph} are trivial. Using the \rf{su2generators} and \rf{wsu2} it follows that the components $\tilde{h}_{\Lambda \,\Lambda}$,  $\tilde{h}_{\Lambda{\cal P}_\pm}$, ${\tilde h}_{{\cal P}_{\pm}\,{\cal P}_{\pm}}$ and ${\tilde h}_{{\cal P}_{\pm}\,{\cal P}_{\mp}}$, given in \rf{hl}, \rf{ppf}, \rf{ppm} and \rf{mmf}, becomes
\br
{\tilde h}_{\Lambda \,\Lambda}&=&-\eta\,\frac{4\,\sin^2\(\frac{f}{2}\)}{\lambda\, f'\,r^2}\, \frac{\(1+\mid z \mid^2\)^2}{\(1+\mid u\mid^2\)^2}\,u'\,\bar{u}';\qquad\qquad \quad {\tilde h}_{T_{+}\,T_{+}}={\tilde h}_{T_{-}\,T_{-}}=-\eta\,\frac{f'}{\lambda}\,\qquad \\
{\tilde h}_{\Lambda\,T_{\pm}}&=& {\tilde h}_{T_{\pm}\,\Lambda}= {\tilde h}_{T_{+}\,T_{-}}={\tilde h}_{T_{-}\,T_{+}}=0
\er
Therefore, the $\tilde{h}$-fields form the diagonal matrix
\be \tilde{h}=-\eta\,\frac{f'}{\lambda}\,{\rm diag.}\,\(1,\,1,\,\frac{4\,\sin^2\(\frac{f}{2}\)}{r^2\, f^{\prime 2}}\, \frac{\(1+\mid z \mid^2\)^2}{\(1+\mid u\mid^2\)^2}\,u'\,\bar{u}'\) \lab{hdiag}\ee
which is fully determined in terms of the fields $f,\,u,\,\bar{u}$, which remains totally free. Note that due \rf{htildedef} the eigenvalues of the $\tilde{h}$ matrix are the same that eigenvalues of the $h$ matrix, which is non-negative. It so follows from \rf{hdiag} that the profile function $f$ must be a monotonic function and
\be {\rm sign}\,\(f'\,\lambda\)=-\eta \lab{flambda}\ee
which due to \rf{signQ} also implies that ${\rm sign}\,\(f'\,Q\)=-\eta$.

For the function $u (\chi)$ to be a well defined map between two-spheres it has
to be a ratio of two polynomials $p(\chi)$ and $q(\chi)$ without commum roots, i.e. the so-called rational map ansatz \cite{mantonbook,rational0,rational1,rational2}
\be u(\chi) = \frac{p(\chi)}{q(\chi)} \lab{rationalmap}\ee
The topological degree of the $u$ map is equal to the highest degree among the polynomials $p(\chi)$ and $q(\chi)$ and can be written in the integral representation as 
\be{\rm deg}\,u=\int \frac{i\,dz \,d\bar{z}}{2\,\pi\,\(1+\mid z \mid^2\)^2}\,\(\frac{1+\mid z \mid^2}{1+ \mid u\mid^2}\,\left|\frac{d u}{dz}\right|\)^2 ={\rm max}\,\left\{{\rm deg}\,p,\,{\rm deg}\,q\right\}\lab{chargesu2tex}
\ee
Therefore, using \rf{wsu2} and \rf{chargesu2tex} the topological charge \rf{pretopcharge3d} become 
\be Q=\eta\,\frac{\left[f-\sin f\right]^{r=0}_{r=\infty}}{2\,\pi}\,{\rm deg}\,u \lab{chargesu2} \ee

\section{The $SU(N+1)/SU(N)\otimes U(1)$ Hermitian symmetric space}
\label{sec:sun}
\setcounter{equation}{0}

\begin{figure}[htp]
\begin{center}
		\includegraphics[scale=0.5]{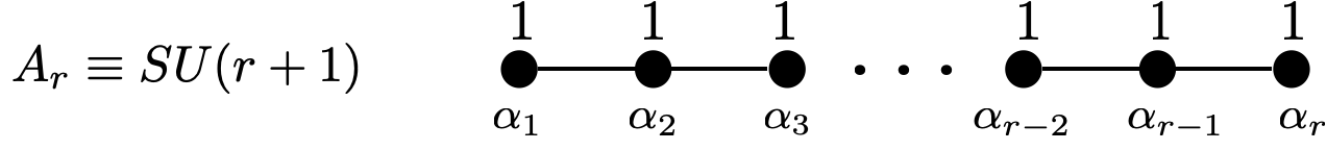}
			\caption{The Dynkin diagrams of the simple Lie algebra $A_r$. The $\alpha_a$'s below the spots label the simple roots, and the numbers above correspond to the integers $m_a$ in the expansion of the highest root $\psi=\sum_{a=1}^r m_a\,\alpha_a$, while the black spots correspond to $m_a=1$.}
		\label{fig:dynkin}
\end{center} 
\end{figure}

For the Hermitian symmetric space $CP^N=SU(N+1)/SU(N)\otimes U(1)$ we choose $\alpha_*=\alpha_N$, which implies $\lambda_*=\lambda_N$, and work with the fundamental $(N+1)\times (N+1)$ representation of $SU(N+1)$ (see Figure \ref{fig:dynkin}). The $S$ matrix is is parametrized by $N$ complex scalar fields $u_a=u_a\(\chi\)$, with $a=1,\,...,\, N$, corresponding with the components of $u^T=\(u_1,\,...,\,u_N\)$. The $\Lambda$ and $S$ matrices defined respectively by \rf{sigmadef}  and \rf{sdef} are given by  
\be
\lab{S3a}
\Lambda=\frac{1}{N+1}\(\begin{array}{cc}
\one_{N \times N} & 0 \\0 & -N
\end{array}\)\qquad\quad {\rm and}\quad \qquad  S =\(\begin{array}{cc} O_{N\times N} & u \\ O_{1\times N} & 0 \end{array} \) 
 \ee
where $O_{1\times N}$ is a $1\times N$ zero matrix, and so on. The $S$ matrix \rf{S3a} satisfies \rf{omegadef} with $\omega= u^\dagger\,u$ and so the $g$ elements \rf{gdef2} which parametrizes the $CP^N$ are given by the unitary matrix
\be g=\frac{1}{\vartheta}\(\begin{array}{cc}
\Delta & i\,u \\i\,u^\dagger & 1
\end{array}\) \lab{mat}
\ee
where $\Delta$ is a $N\times N$ Hermitian matrix defined by 
\be \Delta\equiv \vartheta \,\one_{N \times N}+\(1-\vartheta\)\,T_u\,; \qquad\quad {\rm with } \qquad\quad \vartheta \equiv \sqrt{1+u^\dagger u} \lab{deltadef}\ee
where $T_u\equiv \frac{u \,\otimes \, \bar{u}}{u^\dagger u}$ is a projector, i.e. $T_u^2=T_u$. Note that due to \rf{deltadef} $u$ is an eigenvector of $\Delta$ with eigenvalue $+1$, i.e. $\Delta u = u$, which also implies that $u^\dagger\Delta =u^\dagger$ and $\Delta^{-1}\,u=u$.

The two sets of $N$ abelian matrices $P_+^a\in {\cal P}_+$ and $P_-^a\in {\cal P}_-$ associated with \rf{mat} are given by 
\be \({ P}_+^a\)_{bc} \equiv \delta_{b\,a}\,\delta_{c\,(N+1)} \, ,\qquad\qquad\qquad \({ P}_-^a \)_{bc}= \({P}_+^{a\,\dagger}\)_{bc} = \delta_{b\,(N+1)}\,\delta_{c\,a} \lab{p+-}\ee
with $b,\,c = 1,\,...,\, N+1,\: {\rm and } \:\, a =1,\,...,\, N$.  The two sets of $N$ generators $T^{2\,a-1}$ and $T^{2\,a}$, with $a=1,...,\,N$, are related to $P^a_\pm$ through $P^a_\pm = T^{2\,a-1}\pm i\,T^{2\,a}$, and satisfies the orthogonality relation \rf{kappadef} with $\kappa=\frac{1}{2}$. Therefore, together with \rf{mat} such a generators  satisfies \rf{hermitianniceprop}, and in additon we have 
\be \Tr \,\({ P}_\pm^a\,{ P}_\pm^b\)=0;\qquad\qquad\qquad \Tr \,\({ P}_\pm^a\,{ P}_\mp^b\)=\delta_{ab};\qquad\qquad  \sbr{\Lambda}{P_{\pm}^a}=\pm P_{\pm}^a\lab{tracepmn}\ee
Note that the last equation of \rf{tracepmn} corresponds to \rf{hermitianniceprop}.\footnote{ Clearly, the equation \rf{hermitianniceprop} is invariant by transformations $P_\pm^a \rightarrow \alpha\,P_\pm^a$ with $\alpha$ being any complex number. Therefore, the only matrix proportional to $\Lambda$ that satisfies \rf{hermitianniceprop} is itself, which is given by \rf{mat}.} 

The $N^2-1$ generator of the group ${\cal H} = SU(N)$ can be broken in three set of generators. The first two sets ${\cal H}_R$ and ${\cal H}_I$ contains $\frac{N(N-1)}{2}$ generators each and can be labeled by the pair $nm$, with $n=1,\,...,\,N$ and $m=1,\,...,\,n-1$, i.e. $m<n$.  The third set ${\cal H}_s$ contains $N-1$ generators and is labeled by the index $s=1,\,...,\,N-1$. The generators of such sets can be written respectively, for all $a,\,b=1,\,...,\,N$, as 
\br \({H}_R^{nm}\)_{ab}&=&\frac{1}{2} \,\(\delta_{an}\,\delta_{bm}+\delta_{am}\,\delta_{bn}\)\nonumber\\
\({H}_I^{nm}\)_{ab} &=&-\frac{i}{2}\,\(\delta_{an}\,\delta_{bm}-\delta_{am}\,\delta_{bn} \) \lab{hris}\\
\({H}_{s}\)_{ab} &=&\frac{1}{2}\,\(\delta_{as}\,\delta_{bs}-\delta_{a(s+1)}\,\delta_{b(s+1)}\)\nonumber \er
Note that ${H}_R^{nm},\,{H}_I^{nm}$, and ${H}_{s}$ are extensions of the Pauli matrices $\sigma_1,\,\sigma_2,\,\sigma_3$, respectively, and for such basis $\kappa=\frac{1}{2}$. It so follows that
\be {\rm dim}\,{\cal H} = N^2-1 ;\qquad\qquad {\rm dim}\,{\cal P}=2\,{\rm dim}\,{\cal P}_+ =2\,{\rm dim}\,{\cal P}_- =2\, N \lab{dimension1}\ee
In particular, for the $SU(2)$ case $(N=1)$ we have ${\cal H}=\emptyset$ (see \rf{su2generators}).

Due \rf{mat} and \rf{p+-} the quantities $K_i$ and $P^{(\pm)}_i$ defined in \rf{ws} and \rf{wpm} so becomes
\be
{K}_i=-\frac{\pa_i \vartheta}{\vartheta} \one +\frac{1}{\vartheta^2}\left(\begin{array}{cc} \Delta\, \pa_i \Delta+u \,\otimes\,\pa_i \bar{u} &0 \\0 & u^\dagger \,\pa_i u \end{array}\right) 
\lab{wk}\ee
\be P^{(+)}_i=\frac{i\,\(\Delta\,\pa_i u\)_a }{\vartheta^2}\,P^a_+ \, ; \qquad\qquad\quad P^{(-)}_i=\frac{i\,(\pa_i u^\dagger\,\Delta)_a }{\vartheta^2}\,P^a_- \lab{w+-}\ee
where there is an implicit sum over the index $a$. In addition, the topological charge \rf{pretopcharge3d} becomes 
\be Q =\eta\,\frac{\left[f-\sin f\right]^{r=0}_{r=\infty}}{2\,\pi}\,\int \frac{\mid\Delta\,u'\mid^2}{(1+ \mid u\mid^2)^2}\,\frac{i\,dz \,\wedge\, d\bar{z}}{2\,\pi} \lab{Q1}\ee

Using \rf{S3a}, \rf{p+-}, \rf{hris} and \rf{w+-} the fields $\tilde{h}_{\Lambda\,{\cal P_\pm}}$ and $\tilde{h}_{\Lambda\,{\cal H}}$ fixed through \rf{hl} become
\br
 \tilde{h}_{\Lambda \Lambda} &=& \beta\, \mid \Delta \, u'\mid^2 ; \quad \qquad\tih_{\Lambda \map_+^b}=\tih_{\Lambda \map_-^b}=0 ;\qquad \quad\tih_{\Lambda {H}_s} = \frac{1}{2}\,\beta\,\(M_{ss}-M_{(s+1)(s+1)}\) \nonumber\\
 \tih_{\Lambda {H}_R^{nm}}&=&\frac{1}{2}\,\beta\,\( M_{nm}+M_{mn}\)\; ;\qquad \quad\tih_{\Lambda {H}_I^{nm}}=\frac{1}{2}\,\beta\,i\,\(M_{nm}-M_{mn} \)  
\er 
where 
\be \beta\equiv -\frac{\eta}{\lambda\, f'}\,\frac{(N+1)}{N}\,\frac{2\,\sin^2 \(\frac{f}{2}\)}{r^2}\,\frac{\(1+z\,\bar{z}\)^2}{\(1+u^\dagger \, u\)^2}; \quad\qquad\quad \quad M_{nm}\equiv (\Delta\, u')_n \,(u^{\prime \dagger} \,\Delta)_m \lab{alpham}\ee
On the other hand, the self-dual equations for the fields $\tilde{h}_{{\cal P_\pm}{\cal H}}$ and $\tilde{h}_{{\cal P_\pm}{\cal P_\mp}}$, as given respectively in  \rf{ph} and \rf{ppm}, are reduced to
\be (u^{\prime\dagger}\Delta)_a\,\tih_{P_+^a {\cal H}}=(\Delta u')_a \tih_{P_-^a {\cal H}}=(\Delta\, u')_a\,\tih_{P_-^a P_+^b}=0 \lab{sdph} 
\ee
while the self-dual equations for the fields $\tilde{h}_{{\cal P_+}{\cal P_+}}$ and $\tilde{h}_{{\cal P_-}{\cal P_-}}$, as given respectively in \rf{ppf} and \rf{mmf}, are reduced to
\be (\Delta\, u')_a\, \tih_{P_-^a P_-^b}+\eta\,\frac{f'}{\lambda}\,  (\Delta\,u')_b=( u^{\prime \dagger} \Delta)_a\, \tih_{P_+^a P_+^b}+\eta\,\frac{f'}{\lambda}\,(u^{\prime \dagger} \Delta)_b
=0\lab{sdpp}\ee

Using \rf{dimension1}, the first and second equation of \rf{sdph}, from the left to the right, forms each a set of ${\rm dim}\,{\cal H}=N^2-1$ linear equations for the $N\,(N^2-1)$ fields $\tih_{P_+^a {\cal H}}$ and the $N\,(N^2-1)$ fields $\tih_{P_-^a {\cal H}}$, respectively. Since $\tilde{h}$ is symmetric, the third equation of \rf{sdph}, from the left to the right, forms a set of ${\rm dim}\,{\cal P}_+=N$ linear equations for the $N^2$ fields $\tih_{P_-^a P_+^b}$. Finally, the first and second equation of \rf{sdpp}, from the left to the right, forms each a set of ${\rm dim}\,{\cal P}_+=N$ linear equations for the $N\,(N+1)/2$ fields $\tih_{P_+^a P_+^b}$ and the $N\,(N+1)/2$ fields $\tih_{P_-^a P_-^b}$, respectively.

Consequently, only for $N=1$ we have enough equations for determine such a components of the $\tih$ fields. However, for such case ${\rm dim}\,{\cal H}=0$ and ${\rm dim}\,{\cal P}_+=1$. Thus, although we have four equations and only three independent fields in $\tih_{{\cal P}{\cal P}}$, it so follows that the third and fourth equation of \rf{sdph}, from the left to the right, which corresponds to one equation each, become equivalent. Thus, the $\tih$-fields are totally determined in terms of the fields $f,\,u,\,\bar{u}$, which remains totally free, as we shown in the section \ref{sec:su2}.

\subsection{An explicit example: exact generalized Skyrmions for each integer value of $Q$ on the $CP^N$ spaces}

For construct an explicit self-dual configurations consider the ansatz where all the $u$-fields are equal to the same holomorphic rational map $u_1(\chi)=p(\chi)/q(\chi)$ between the Riemann spheres $S^2$ (see \rf{rationalmap}), the $\tih_{{\cal P}{\cal P}}$-fields forms a diagonal matrix and $\tih_{{\cal P}{\cal H}}=0$, i.e.
\be
u_a= u_1=\frac{p(\chi)}{q(\chi)};\qquad\qquad\quad  \tih_{{\cal P}{\cal H}}=\tih_{{ P}_{\pm}^a{P}_\mp^b}=0;\quad\qquad\qquad  \tih_{{P}_{\pm}^a{ P}_\pm^b}=\delta_{ab}\,\tih_{{ P}_{\pm}^a{ P}_{\pm}^b}\; 
\lab{speciala}\ee 
for $a,\,b=1,\,...,\,N$ and where there is no implicit sum over $a$ or $b$. It so follows that all self-dual equation given in \rf{sdph} are automatically satisfied, while \rf{sdpp} imposes that $\tih_{{\cal P}{\cal P}}$ must be
\be \tih_{\map \map}= -\eta\,\frac{f'}{\lambda} \,\one_{2\,N\times 2\,N}\lab{hpp2}\ee
On the other hand, using the ansatz \rf{speciala} and the definitions \rf{deltadef} and \rf{alpham} we obtain $(\Delta\,u')_n = u_1'$, with $n=1,\,...,\,N$, which due to \rf{speciala} implies $M_{nm}=u_1'\bar{u}_1'$. Therefore, the field given in \rf{ppm} are reduce to
\begin{align}
\lab{hh2}
\begin{split}
\tih_{\Lambda\Lambda}&=-\frac{\eta}{\lambda\, f'}\,\frac{2\,(N+1)\,\sin^2 \(\frac{f}{2}\)}{r^2}\,\frac{\(1+z\,\bar{z}\)^2}{\(1+N\,\mid u_1 \mid^2\)^2}\,\bar{u}_1'\, u'_1\\
\tih_{\Lambda {\cal P}_\pm}&=\tih_{\Lambda {\cal H}_I}=\tih_{\Lambda {\cal H}_s}=0;\qquad \quad\tih_{\Lambda \mathcal{H}_R^{nm}}=\frac{1}{N}\,\tih_{\Lambda\Lambda}
\end{split} 
\end{align}
where we used $u^{\prime \dagger}\Delta^2 u'=\sum_{n=1}^{N} M_{nn}= N\,\bar{u}_1'\,u_1'$. Therefore, the only non-vanishing components of $\tih_{\Lambda {\cal H}}$ are the $\tih_{\Lambda {\cal H}_R}$-fields, which in turns forms a column with all components equal to $N^{-1}\tih_{\Lambda\Lambda}$. An interesting consequence is that the non-singular $\tilde{h}$-matrix inside the ansatz \rf{speciala} is non-diagonal for $N>1$. On the other hand, the field configuration \rf{speciala}-\rf{hh2} is a clear generalization of the \rf{hdiag}, as obtained for $N=1$ in the section \ref{sec:su2}. In fact, the $\tilde{h}$ matrix has the explicit form
\begin{table}[H]
\centering
\begin{tabular}{c|c|c|c|c|c|c|}
& $\mathcal{H}_R$ & $\mathcal{H}_I$ &$\mathcal{H}_s$ & $\map_+$ & $\map_-$ & $\Lambda$\\
\hline                              
$\mathcal{H}_R$ & & & & 0& 0& $\tih_{\mathcal{H}_R \Lambda}$  \\ \hline
 $\mathcal{H}_I$ & & & & 0&0 & 0 \\ \hline
$\mathcal{H}_s$  & & & & 0& 0& 0 \\ \hline
$\map_+$ & 0& 0& 0& $\tih_{\map_+ \map_+}$& 0& 0 \\ \hline
$\map_-$ & 0& 0& 0& 0&  $\tih_{\map_+ \map_+}$&  0 \\ \hline
$\Lambda$ & $\tih_{\Lambda \mathcal{H}_R}$ & 0&0 &0 &0 & $\tih_{\Lambda \Lambda}$  \\ \hline 
\end{tabular}
\end{table}
\noindent where the blank spaces are the free $\tilde{h}_{{\cal H}{\cal H}}$ components, the zeros clearly are null matrices, $\tih_{\map_\pm \map_\pm}= -\eta\,\lambda^{-1}\,f' \,\one_{N\times N}$, and $\tih_{\mathcal{H}_R \Lambda}$ is a ${\rm dim}\,\mathcal{H}_R \times 1$ matrix with all the components equal to $N^{-1}\,\tilde{h}_{\Lambda\Lambda}$. On the other hand, the topological charge \rf{Q1} is reduced to 
\be Q =\eta\,\frac{\left[f-\sin f\right]^{r=0}_{r=\infty}}{2\,\pi}\,{\rm deg}\,u_1 \lab{chargeex0} \ee

Let us consider the boundary conditions for the perfil function $f\(0\)= 2\,\pi\,m$  and $f\(\infty\)= 0$ for some sign function $\eta'=1$, and $f\(0\)= 0$  and $f\(\infty\)= 2\,\pi\,m$ for $\eta'=-1$, where $m$ is any positive integer. Clearly, these boundary conditions ensures that $Q$ is an integer. The algebraic degree of the rational map, which corresponds to the the highest degree among the polynomials $p(\chi)$ and $q(\chi)$, denoted by a positive integer $n$, is equal to the topological degree of the map $u_1(\chi)$. Thus, the topological charge \rf{chargeex0} inside the rational map ansatz become
\br Q = \eta'\,\eta\,m\,n \lab{chargeex1}
\er
which due to \rf{signQ} implies ${\rm sign}\(\lambda\)=\eta'\,\eta$.

As the rational map and the profile function are still free, we have an infinite number of exact solutions for any integer value of the topological charge and for each value of $N$. By example, for the Hermitian symmetric space $SU(3)/SU(2)\otimes U(1)$ $(N=2)$ let us consider the radial solutions  $p(\chi)=\chi$ and $q(\chi)=\sqrt{N}$, i.e. $u_1=\frac{\chi}{\sqrt{N}}$, which turn the topological charge and static energy densities, as well the $\tilde{h}$-fields, spherically symmetric. In such a case $\dim {\cal H}_s=\dim {\cal H}_R=\dim {\cal H}_I=1$, $\dim {\cal P}_+=\dim {\cal P}_-=2$ and the $\tilde{h}$ matrix become
\be
\lab{hex1}
\tilde{h}=  -\eta' \,f' \, \(\begin{array}{cccccccc}
 & & & 0 & 0 & 0 & 0 &  \frac{3\,\sin^2 \(\frac{f}{2}\)}{2\,f^{\prime 2}\,\zeta^2}\\
 & -\frac{\eta'}{f'} \,\tilde{h}_{{\cal H}{\cal H}} &  & 0 & 0 & 0 & 0 & 0 \\
 & & & 0 & 0 & 0 & 0 & 0 \\
0 & 0 & 0 & 1 & 0 & 0 & 0 & 0 \\ 
0 & 0 & 0 & 0 & 1 & 0 & 0 & 0 \\ 
0 & 0 & 0 & 0 & 0 & 1 & 0 & 0 \\ 
0 & 0 & 0 & 0 & 0 & 0 & 1 & 0 \\ 
\frac{3\,\sin^2 \(\frac{f}{2}\)}{2\,f^{\prime 2}\,\zeta^2} & 0 & 0 & 0 & 0 & 0 & 0 & \frac{3\,\sin^2 \(\frac{f}{2}\)}{f^{\prime 2}\,\zeta^2} \\     
\end{array}\)
\ee
where we introduce the dimensionless radius $\zeta \equiv \mid \lambda \mid \,r$, and we use  $\lambda = \eta'\,\eta\,\mid \lambda\mid$.  

Now, let us choose $\tilde{h}_{{\cal H}{\cal H}}=-\eta'\,f'\,\one_{3\times 3}$ and take the example $f=4\,m\,\arctan\(\(\frac{a}{\zeta}\)^{\eta'}\)$, where $a$ is and arbitrary positive dimensionless constant. This choice of the perfil function and the $\tilde{h}_{{\cal H}{\cal H}}$ terms preserves the positivity of the $\tilde{h}$ matrix, reducing \rf{hex1} to the spherically symmetric form
\be
\tilde{h}=  \delta  \, \(\begin{array}{ccc}
 1 & O_{1\times 6} & \frac{1}{2}\,\gamma\\
O_{6\times 1} & \one_{6\times 6}& O_{6\times 1} \\
 \frac{1}{2}\,\gamma & O_{1\times 6} &  \gamma \\     
\end{array}\) 
\ee
with $\gamma \equiv \frac{3\,\sin^2 \(\frac{f}{2}\)}{f^{\prime 2}\,\zeta^2} = \frac{\(a^2+\zeta^2\)^2}{4\, a^2\,m^2\,\zeta^2}\,\sin^2\(2\,m\,\arctan\(\(\frac{a}{\zeta}\)^\eta\)\)$, $\delta \equiv \frac{4\,a\,m\,\mid \lambda \mid}{a^2	+\zeta^2}$, and where $O_{1\times 6}$ denotes a $1\times 6$ zero matrix, and so on. The $\tilde{h}$ matrix have six eigenstates equal to $\delta$ and the other two are $\frac{1}{2}\,\delta\,\(1+\gamma \pm \sqrt{1+2\,\( -1+\gamma)\,\gamma\)}\)$. However, as we proof in the Appendix \ref{sec:appendix2}, for such a perfil function we have 
\be  g\(\zeta\)\equiv \frac{4\,\sin^2 \(\frac{f}{2}\)}{f^{\prime 2}\,\zeta^2} =\frac{\(a^2+\zeta^2\)^2}{4\,m^2\,a^2\,\zeta^2}\sin^2 b \leq 1 ;\qquad\qquad b\equiv 2\,m\,\arctan\(\(\frac{a}{\zeta}\)^{\eta'}\) \lab{fine}\ee
which implies $0 \leq \gamma =\frac{3}{4}\,g \leq g  \leq 1$. Therefore, all eigenvalues of the $\tilde{h}$ matrix are non-negative. In particular, for the $Q=\eta'\,\eta$ topological solutions $(m=1)$, we have $\gamma=\frac{3}{4}$ and $\delta= \frac{4\,a\,\mid \lambda \mid}{a^2	+\zeta^2}$ and therefore the diagonal components of $\tilde{h}$ are equal and all non-vanishing $\tilde{h}$ fields fall asymptotically with $1/\zeta^2$.

\section{The case of $SU\(p+q\)/SU\(p\)\otimes SU\(q\)\otimes U\(1\)$} \label{Spq}
\setcounter{equation}{0}

In the case of the Hermimtian symmetric space $SU\(p+q\)/SU\(p\)\otimes SU\(q\)\otimes U\(1\)$, we choose $\alpha_*=\alpha_p$, which implies $\lambda_*=\lambda_p$, the fundamental $(p+q)\times (p+q)$ representation of $SU(p+q)$. The $S$ matrix is parametrized by $p$ complex scalar fields $u_a=u_a(\chi)$, with $a=1,\,...,\,p$, and $q$  complex scalar fields $v_b=v_b(\chi)$, with $b=1,\,...,\,q$, corresponding with the components of $u^T=\(u_1,\,...,\,u_p\)$ and $v^T=\(v_1,\,...,\,v_q\)$. The $\Lambda$ and $S$ matrices defined respectively by \rf{sigmadef}  and \rf{sdef} are given by  
\be
\Lambda = \frac{1}{p+q}\,\(\begin{array}{cc} q\,\one_{p\times p} & O_{p\times q}\\ O_{q\times p} & -p\,\one_{q\times q}\end{array}\)\qquad\quad {\rm and}\quad \qquad  S = \(\begin{array}{cc} O_{p\times p} & u \, \otimes\, v\\ O_{q\times p} & O_{q\times q}\end{array} \) \lab{Spq}
\ee
where $O_{p\times q}$ is a $p\times q$ zero matrix, and so on. We consider that both the fields $u$ and $v$ are (anti)-holomorphic when $S$ is (anti)-holomorphic.  The $S$ matrix \rf{S3a} satisfies \rf{omegadef} with $\omega= \mid u \mid^2\,\mid v \mid^2$  and so the $g$ elements given in \rf{gdef2} become
\be
\lab{gpq} g=\frac{1}{\vartheta}\,\(\begin{array}{cc} \Delta_u & i\,u \, \otimes\, v \\ i\,\overline{v}\,\otimes\, \bar{u} & \Delta_v^T \end{array}  \)\,;\qquad\qquad\qquad \Delta_x \equiv \vartheta \,\one + \(1-\vartheta\)\, T_x 
\ee
with $\vartheta = \sqrt{1 + \omega}$, and where $T_x\equiv \frac{x\,\otimes\, \bar{x}}{x^\dagger x}$ is a projector, i.e. $T_x^2=T_x$ and $x$ is any complex vector. The operator $\Delta_x$ is Hermitian and inversible, its inverse corresponds to $\Delta_x^{-1} = \vartheta^{-1}\,\( \one - \(1-\vartheta\)\, T_x\) $ and its square to $\Delta_x^2=(1+\omega)\,\one-\omega\,T_x$. The vetor $x$ is an eigenvector with eigenvalue $+1$ of both operators $\Delta_x$ and $\Delta_x^{-1}$. It so follows that $\Delta u = u$,  $\Delta_v^T\,\bar{v}= \bar{v}$, $\Delta_v^T\,\bar{v}= \bar{v}$ and $u^\dagger\,\Delta_u=u^\dagger$. 

The $\Lambda$ matrix \rf{Spq} satisfies $\Lambda^2=c\,\Lambda + \frac{1}{4}\,\(1-c^2\)$ with $c=\frac{q-p}{q+p}$, then the field $U$ have the form given in \rf{Uexpanded} (see section \ref{sec:holomorphic}), i.e. 
\be U=e^{\frac{i\,q\,f(r)}{p+q}}\,\left[\one +\(e^{-i\,f(r)}-1\)\,Z\right]\,;\qquad {\rm with}\qquad Z= \frac{1}{\vartheta^2}\,\(\begin{array}{cc} \omega\, T_u & i\,u\,\otimes \, v \\-i\,\bar{v}\,\otimes \,\bar{u} & \Delta_v^{T\,2} \end{array}\) \lab{upq}\ee 
The reduction to the $CP^N$ case occurs by imposing  $p=N$, $q=1$ with $v=v_1$ implying $\Delta_v=1$. In addition, the the field $v_1$ can be absorbed in the field $u$ through the transformation $u\rightarrow u/v_1$, which reduces $\Delta_u$ to \rf{deltadef} and $g$ to \rf{mat}. This transformation is equivalent to setting $v=1$. 

The two sets of $p\times q$ abelian generators $P_+^{cd}$ and $P_-^{cd}$, with $c=1,\,...,\,p$ and $d=1,\,...,\,q$, of the ${\cal P}_+$ and ${\cal P}_-$ subalgebras are given by 
\be \({ P}_+^{cd}\)_{ab} \equiv \delta_{a\,c}\,\delta_{(d+p)\,b} \, ,\qquad\qquad\qquad \({ P}_-^{cd} \)_{ab} = \delta_{a\,(d+p)}\,\delta_{b\,c} \lab{psupq}\ee 
with $a,\,b = 1,\,...,\, p+q$, which satisfies  $\Tr\({P}_+^{cd}\,{ P}_-^{c'd'}\)=\delta_{cc'}\,\delta_{dd'}$, with $c'=1,\,...,\,p$ and $d'=1,\,...,\,q$. There are $p^2-1$ generators of the group $SU(p)$ and $q^2-1$ generators of the group $SU(q)$ associated with the $H=SU(p)\otimes SU(q)$ subgroup of $SU(p+q)$. For each of such a groups we can break such a generators in tree distinct types, similarly that we do in \rf{hris}. In case of the $SU(p)$ group, the first two sets ${\cal H}_{R_p}$ and ${\cal H}_{I_p}$ contains $\frac{p\,(p-1)}{2}$ generators each and can be labeled by the pair $nm$, with $n=1,\,...,\,p$ and $m=1,\,...,\,p-1$, i.e. $p<n$.  The third set ${\cal H}_{s_p}$ contains $p-1$ generators and is labelled by the index $s=1,\,...,\,p-1$. The same follows for the $SU(q)$ group by changing $p\rightarrow q$ and changing the indices $m\rightarrow k$, $n\rightarrow l$ and $s\rightarrow r$. The generators of the $SU(p)$ and $SU(q)$ groups are given by
\br \({H}_{R_p}^{nm}\)_{ab}&=&\frac{1}{2} \,\(\delta_{an}\,\delta_{bm}+\delta_{am}\,\delta_{bn}\) \,;\qquad \;\;\({H}_{R_q}^{lk}\)_{ab}=\frac{1}{2} \,\(\delta_{a\,(l+p)}\,\delta_{b\,(k+p)}+\delta_{a\,(k+p)}\,\delta_{b\,(l+p)}\) \nonumber\\
\({H}_{I_p}^{nm}\)_{ab} &=&-\frac{i}{2}\,\(\delta_{an}\,\delta_{bm}-\delta_{am}\,\delta_{bn} \) \,;\quad\;\;\; \({H}_{I_q}^{lk}\)_{ab} =-\frac{i}{2}\,\(\delta_{a\,(l+p)}\,\delta_{b\,(k+p)}-\delta_{a\,(k+p)}\,\delta_{b\,(l+p)} \) \nonumber\\
\({H}_{s}\)_{ab} &=&\frac{1}{2}\,\(\delta_{as}\,\delta_{bs}-\delta_{a(s+1)}\,\delta_{b(s+1)}\)\,;\;\; \({H}_{r}\)_{ab} =\frac{1}{2}\,\(\delta_{a\,(r+p)}\,\delta_{b\,(r+p)}-\delta_{a\,(r+p+1)}\,\delta_{b\,(r+p+1)}\) \nonumber\\
 {\rm with } & &a,\,b=1,\,...,\,p+q  \lab{hrisp} \qquad\er
Note that for such a basis we have 
\be \kappa=\frac{1}{2} \lab{kappapq}\ee

Using \rf{omegas} and \rf{gpq} we get
\be
\Omega^{-1}= \frac{1}{\vartheta}\,\(\begin{array}{cc} \Delta_u &  O_{p\times q}\\ O_{q\times p} & \Delta_v^T\end{array}\) 
\ee
which together with \rf{extraction} fixes $P_\chi^{(+)} = i \,\Omega^{-1}\,\pa_\chi S\,\Omega^{-1}$,  $P_{\bar{\chi}}^{(-)} = i \,\Omega^{-1}\,\pa_{\bar{\chi}} S^\dagger\,\Omega^{-1}$ and the commutator $\left[P_{\chi}^{(+)},\,P_{\bar{\chi}}^{(-)}\right]$ through 
\br 
P_\chi^{(+)}& =& i\,\vartheta^{-2}\,\(\begin{array}{cc} O_{p\times p} & B\\ O_{q\times p} & O_{q\times q}\end{array}\)\,; \qquad\,\,\,\,  P_{\bar{\chi}}^{(-)} = i\,\vartheta^{-2}\,\(\begin{array}{cc} O_{p\times p} & O_{p\times q}\\ B^\dagger &O_{q\times q}\end{array}\) \nonumber\\
\left[P_{\chi}^{(+)},\,P_{\bar{\chi}}^{(-)}\right] &=& -\frac{1}{\vartheta^4}\(\begin{array}{cc} B\,B^\dagger & O_{p\times q}\\ O_{q\times p}\ & -B^\dagger\,B \end{array}\) ;\qquad {\rm with }\qquad B\equiv \Delta_u \,\pa_\chi\(u\,\otimes \,v\)\,\Delta_v^T \qquad \lab{comutpp}
\er
From \rf{psupq} and \rf{comutpp} we can also write
\be P_\chi^{(+)} = i\,\vartheta^{-2}\,B_{cd}\, P_+^{cd}\,;\qquad\qquad\qquad P_{\bar{\chi}}^{(-)} = - \(P_\chi^{(+)} \)^\dagger= i\,\vartheta^{-2}\,B_{dc}^\dagger\, P_-^{cd} \lab{psb}\ee
On the other hand, using $\omega=\vartheta^2-1$, the definition of the operators $\Delta_x$ and $T_x$ introduced in \rf{gpq} we obtain
\be
B = \vartheta\,\left[\pa_\chi \(u\,\otimes \,v\) -  \frac{2\,\pa_\chi \vartheta}{1+\vartheta} \,u\,\otimes \,v \right]\,;\qquad \quad
B^\dagger = \vartheta\,\left[\pa_{\bar{\chi}} \(\bar{v}\,\otimes \,\bar{u}\) -  \frac{2\,\pa_{\bar{\chi}} \vartheta}{1+\vartheta} \,\bar{v}\,\otimes \,\bar{u} \right] \lab{bbd}
\ee

Using \rf{Spq}, \rf{psupq}-\rf{bbd} the fields $\tilde{h}_{\Lambda\,{\cal P_\pm}}$ and $\tilde{h}_{\Lambda\,{\cal H}}$ fixed through \rf{hl} become
\br  \tilde{h}_{\Lambda \Lambda} &=& \beta\,\Tr\(M\)  ;\,\,\qquad\qquad\qquad\qquad \qquad\tih_{\Lambda \map_\pm}=0 \nonumber\\
\tih_{\Lambda {H}_s} &=& \frac{1}{2}\,\beta\,\(M_{ss}-M_{(s+1)(s+1)}\) \,;\qquad\quad \quad \tih_{\Lambda {H}_r} = \frac{1}{2}\,\beta\,\(N_{rr}-N_{(r+1)\,(r+1)}\)\nonumber\\
\tih_{\Lambda {H}_{R_p}^{nm}}&=&\frac{1}{2}\,\beta\,\( M_{nm}+M_{mn}\) \,;\qquad\qquad \qquad \,\,\tih_{\Lambda {H}_{R_q}^{lk}}=\frac{1}{2}\,\beta\,\(N_{lk}+N_{kl}\)\nonumber\\
\tih_{\Lambda {H}_{I_p}^{nm}}&=&\frac{1}{2}\,\beta\,i\,\(M_{nm}-M_{mn} \)   \,;\qquad\qquad \qquad \tih_{\Lambda {H}_{I_q}^{lk}}=\frac{1}{2}\,\beta\,i\,\(N_{lk}-N_{kl} \)  \lab{hlambdapq}
\er 
where 
\be \beta\equiv -\eta\,\vartheta^{-4}\,\alpha= -\eta \,\frac{p+q}{q\,p}\,\frac{2\,\sin^2\frac{f}{2}}{\lambda\,f' \,r^2}\,\frac{\(1+\mid z\mid^2\)^2}{\vartheta^4} \,;\qquad M\equiv B\,B^\dagger\,; \qquad N\equiv B^\dagger\,B \lab{betasq}
\ee
and the explicit form  of $M$, $N$ and the trace $\Tr\(M\)=\Tr\(N\)$ corresponds to 
\br
M &=&  \vartheta^2\,\left\{ \pa_\chi \pa_{\bar{\chi}} \(\omega \, T_u\) -  \frac{2}{1+\vartheta}\,\left[\(\pa_\chi \vartheta\,\pa_{\bar{\chi}} +\pa_{\bar{\chi}} \vartheta\,\pa_\chi\)\(\omega\, T_u\) \right] + 4\,\frac{\pa_\chi \vartheta\,\pa_{\bar{\chi}} \vartheta}{\(1+\vartheta\)^2}\,\(\omega \, T_u\) \right\}\nonumber \\
N &=&  \vartheta^2\,\left\{\pa_\chi \pa_{\bar{\chi}} \(\omega \, T_v^T\) -  \frac{2}{1+\vartheta}\,\left[\(\pa_\chi \vartheta\,\pa_{\bar{\chi}} +\pa_{\bar{\chi}} \vartheta\,\pa_\chi\)\(\omega\, T_v^T\) \right] + 4\,\frac{\pa_\chi \vartheta\,\pa_{\bar{\chi}} \vartheta}{\(1+\vartheta\)^2}\,\(\omega \, T_v^T\) \right\}\nonumber\\
\Tr\(M\) & = &  2\,\vartheta^2\,\(\vartheta\,\,\pa_\chi\pa_{\bar{\chi}} \vartheta -\pa_\chi \vartheta\,\pa_{\bar{\chi}} \vartheta \) = \vartheta^2\,\left[\pa_\chi \pa_{\bar{\chi}} \omega -\frac{\pa_\chi \omega\,\pa_{\bar{\chi}} \omega}{1+\omega}\right]\lab{mn}
\er
On the other hand, the self-dual equations for the fields $\tilde{h}_{{\cal P_\pm}{\cal H}}$ and $\tilde{h}_{{\cal P_\pm}{\cal P_\mp}}$, as given respectively in  \rf{ph} and \rf{ppm}, are reduced to
\be B_{cd}\,{\tilde h}_{{P}_-^{cd}\,{\cal H}}= B_{dc}^\dagger\,{\tilde h}_{{P}_+^{cd}\,{\cal H}}= B_{cd}\,{\tilde h}_{{P}_-^{cd}\,{\cal P}_+}=0 \ee 
while the self-dual equations for the fields $\tilde{h}_{{\cal P_+}{\cal P_+}}$ and $\tilde{h}_{{\cal P_-}{\cal P_-}}$, as given respectively in \rf{ppf} and \rf{mmf}, are reduced to
\be B_{dc}^\dagger\,{\tilde h}_{{P}_+^{cd}\,{P}_+^{nm}} +\eta\,\lambda^{-1}\,f' \,B_{mn}^\dagger = B_{cd}\,{\tilde h}_{{P}_-^{cd}\,{P}_-^{nm}} +\eta\,\lambda^{-1}\,f' \,B_{nm} = 0  \ee 
Using \rf{comutpp} and \rf{kappapq}, which implies ${\rm Tr}\(P^{(+)}_\chi\,P^{(-)}_{{\bar \chi}}\) = -\vartheta^{-4} \Tr M$, and \rf{mn}, the topological charge \rf{pretopcharge3d} becomes
\br
 Q&=& \frac{1}{2\,\pi}\,\left[f\(r\)-\sin f\(r\)\right]_{r=0}^{r=\infty} Q_{{\rm top}} \nonumber
 \\ Q_{{\rm top}}&=& -\frac{i\,\eta}{2\,\pi}\, \int \frac{dz\,d\bar{z}}{\vartheta^4}\;\Tr M=-\frac{i\,\eta}{2\,\pi}\, \int \frac{dz\,d\bar{z}}{\vartheta^2}\;\left[\pa_\chi \pa_{\bar{\chi}} \omega -\frac{\pa_\chi \omega\,\pa_{\bar{\chi}} \omega}{1+\omega}\right]
 \lab{pretopcharge3db}
\er

\subsection{An explicit example: exact generalized Skyrmions on the $SU\(p+q\)/SU\(p\)\otimes SU\(q\)\otimes U\(1\)$ spaces}

For construct explicit self-dual configurations consider the ansatz where all the components fields $u$ are $v$ are equal to the same holomorphic rational map $u_1(\chi)=p_u(\chi)/q_u(\chi)$ and $v_1(\chi)=p_v(\chi)/q_v(\chi)$, respectively, between the Riemann spheres $S^2$ (see \rf{rationalmap}). By definition, $p_t$ and $q_t$, with $t=u,\,v$, does not share any common root, since $u_1$ and $v_1$ are rational maps. However, we also impose that $p_u$ and $q_v$ does not any share common root, and we impose the same restriction to $p_v$ and $q_u$.  Therefore, the product $u_1\,v_1$ is also a rational map. The ansatz for the fields $u$, $v$, $\tih_{{\cal P}{\cal H}}$ and $\tih_{{\cal P}{\cal P}}$ is an generalization of \rf{speciala} and \rf{hpp2}, and is given by
\br
\nonumber u_c&=& u_1=\frac{p_u(\chi)}{q_u(\chi)};\qquad\qquad\quad  v_d= v_1=\frac{p_v(\chi)}{q_v(\chi)}; \nonumber\\
\tih_{{\cal P}{\cal H}}&=& \tih_{{\cal P}_{\pm}{\cal P}_\mp}=0;\quad\qquad\qquad  \tih_{{P}_\pm^{cd}\,{P}_\pm^{nm}}=-\eta\,\frac{f'}{\lambda} \,\delta_{cn}\,\delta_{dm}\; 
\lab{specialaa}\er 
where ${P}_\pm^{cd}$ are defined in \rf{psupq} and $c,\,n=1,\,...,\,p$ and $d,\,m=1,\,...,\,q$. For this indices, the ansatz \rf{specialaa} implies $\(T_u\)_{an}=p^{-1}$ and $\(T_v\)_{dm}=q^{-1}$, leading due to \rf{mn} to $M_{cn}=M_{11}$ and $N_{dm}=N_{11}$,  where 
\be M_{11} =  \frac{\Tr M}{p}\;\qquad\qquad N_{11} =  \frac{\Tr M}{q}\,;\qquad\qquad \Tr\(M\)=p\,q\,\left| \frac{d}{d\chi}\(u_1\,v_1\)\right|^2 \,\geq \, 0 \ee
Therefore, the fields given in \rf{hlambdapq} become
\br  \tilde{h}_{\Lambda \Lambda} &=& \beta\, \Tr M \,;\qquad\qquad\,\,\,  \tih_{\Lambda {H}_{R_p}^{nm}}=\beta\,M_{11} \,; \qquad\qquad \,\,\,\tih_{\Lambda {H}_{R_q}^{lk}}=\beta\,N_{11} \nonumber\\
\tih_{\Lambda \map_\pm}&=&\tih_{\Lambda {H}_s} = \tih_{\Lambda {H}_r} =\tih_{\Lambda {H}_{I_p}^{nm}}=\tih_{\Lambda {H}_{I_q}^{nm}}= 0 \lab{hlambdapqf}
\er 
while all the self-dual equations given in \rf{sdph} and \rf{sdpp} are automatically satisfied by \rf{specialaa}. Note that using \rf{betasq} all the non-vanishing components of $\tih$ other than the free terms $\tih_{{\cal H}{\cal H}}$ are non-negative if we impose the follow condition over the perfil function $f$
\be -\eta\,{\rm sign}\, \(\frac{f'}{\lambda}\) \geq 0 \qquad\qquad \Rightarrow \qquad\qquad \beta \geq 0 \lab{betapq}
\ee

Using $\omega = p\,q\,\mid u_1\mid^2 \,\mid v_1\mid^2$, the topological charge \rf{pretopcharge3db} become
\br
 Q&=& \left[\frac{f\(r\)-\sin f\(r\)}{2\,\pi}\right]_{r=0}^{r=\infty} Q_{{\rm top}} \,;\qquad\quad  Q_{{\rm top}} = -\frac{i\,\eta}{2\,\pi}\, \int \,\frac{dz\,d\bar{z}\,\left| \frac{d}{d\chi}\(\sqrt{p\,q}\,u_1\,v_1\)\right|^2}{\(1+\mid \sqrt{p\,q}\,u_1\,v_1\mid^2\)^2} \qquad \,\, \lab{chargepq2}
\er
However, the equation \rf{chargesu2tex} shows the integral representation of the degree of a rational map $u$. Such a degree is in particular invariant by a the multiplication $u\rightarrow c \,u$, $\forall c\in \IR_+^*$. Therefore, $Q_{{\rm top}}$ given \rf{chargepq2} corresponds with the integral representation of the rational map $u_1\,v_1$, i.e. $Q_{{\rm top}} = {\rm deg}\,\(u_1\,v_1\)$ and the topological charge becomes \rf{chargepq2}
\be
 Q= \eta\,\left[\frac{f\(r\)-\sin f\(r\)}{2\,\pi}\right]_{r=0}^{r=\infty} \,{\rm deg}\,\(u_1\,v_1\)
 \lab{chargepqend}
\ee
Consequently, by choosing the degree of the rational map $u_1\,v_1$ and the boundary conditions of the perfil function $f$ we get an infinite number of exact self-dual solutions, given by \rf{specialaa} and \rf{hlambdapqf}, for the Hermitian symmetric space $SU\(p+q\)/SU\(p\)\otimes SU\(q\)\otimes U\(1\)$. The only restriction is to choose such a fields and the free term $\tih_{{\cal H}{\cal H}}$ such as that $\tih$ is non-negative, as it is  done in the example given in section \ref{sec:sun}.

\section{Conclusion}
\label{sec:conclusion}
\setcounter{equation}{0}

In the self-dual sector of our generalization of the BPS Skyrme model for any compact Lie group $G$ that leads to a Hermitian symmetric space, our holomorphic ansatz shows that the full determination of the $h$ fields in terms of the Skyrme fields happens only for some particular Lie groups. Although this characteristic of the BPS Skyrmions for the $G=SU(2)$ case is not a general feature of the generalized theory, this model possesses the main symmetries of the original BPS Skyrme model. 

As in the original BPS Skyrme model, the $h$ fields in our generalized BPS Skyrme model  continue to play the same role as the Wess-Zumino term  with respect to breaking the invariance by the parity and target space parity transformations $P$ and $P_g$, respectively, while preserving the symmetry by the composition $P\,P_g$. These properties may shed light in the physical nature of the $h$ fields, which may be related to the chiral anomaly.

Our holomorphic ansatz simplifies drastically the self-dual equations. It leads directly to the determination of the components of $\tilde{h}_{\Lambda\Lambda},\,\tilde{h}_{\Lambda {\cal P}_\pm}, \tilde{h}_{\Lambda {\cal H}}$ in terms of the Skyrme field, and leads to algebraic equations for the $\tilde{h}_{{\cal H} {\cal P}_\pm}, \tilde{h}_{{\cal P}_\pm{\cal P}_\pm}, \tilde{h}_{{\cal P}_\pm{\cal P}_\mp}$ components. However, there are at least a number of  ${\rm dim}\,{\cal P}_+\,\(2\,{\rm dim}\,{\cal P}_+-3\)$ components of $\tilde{h}_{{\cal P}{\cal P}}$ and  $2\,{\rm dim}\,{\cal H}\,({\rm dim}\,{\cal P}_+-1)$ components of $\tilde{h}_{{\cal H}{\cal P}}$ totally free.  Clearly, the freedom of the system grows with the dimension of Lie algebra ${\cal G}$. In fact, the $\tilde{h}$ fields can be entirely determined in term of the Skyrme field inside the holomorphic ansatz \rf{holog} only if ${\cal H}=\emptyset$ and ${\rm dim}\,{\cal P}_+=1$, which corresponds to $G=SU(2)$.

The generalized holomorphic ansatz for $G=SU(N+1)$ leads to an infinite number of exact BPS Skyrmions for all integer values of the topological charge and for all $N\geq 1$. We also show how to construct a more restrictive ansatz based on the usual rational map $S^2\rightarrow S^2$, which fixes all  components of the $\tilde{h}$ matrix except $\tilde{h}_{{\cal H}{\cal H}}$. Using this approach, we gave an example of $\tilde{h}$ matrix that leads to exact spherically symmetric BPS Skyrmions for all integer values of $Q$ and $N$. The self-dual sector within the holomorphic ansatz for the Hermitian symmetric space $SU\(p+q\)/SU\(p\)\otimes SU\(q\)\otimes U\(1\)$ is quite similar to the $CP^N$ case, dispite being a generalization. In fact, we can even obtain particular solutions for each value of the topological charge, where  all the non-diagonal entries of the $\tih$ matrix vanish, expect the terms $\tih_{\Lambda {\cal H}_{R_p}^{nm}}$ and $\tih_{\Lambda {\cal H}_{R_q}^{nm}}$.

Our theory facilitates the construction of highly symmetric multi BPS Skyrmions, and extensions of this model may have some important physical applications. One example is the generalization of the False Vacuum Skyrme model to $G=SU(N+1)$. This is very promissing since such a theory is strongly based on spherical symmetric multi-solitions, which also appears in our generalized BPS Skyrme model. On the other hand, our holomorphic ansatz for Hermitian symmetric spaces may be of great value in constructing multi-solitons in a vast number of similar theories.
 
Extensions of the generalized BPS Skyrme model may break both the self-duality equations and the conformal invariance in three spatial dimensions. This can be achieved by introducing kinetic and potential terms for the $h$ fields into the action, as done in the Quasi-Self-Dual model proposed in \cite{quasi-self-dual} for $G = SU(2)$. This may result in the full determination of all fields of the model,  as is the case in \cite{quasi-self-dual}. Another application of our work is the construction of a generalization of the $SU(2)$ False Vacuum Skyrme model introduced in \cite{luiz:false} to larger groups. These applications may shed light on the physical meaning of the $h$ fields, which could depend on the type of extension. 
 
\vspace{2cm}

{\bf Acknowledgements:} LAF is supported by Conselho Nacional de Desenvolvimento Cient\'ifico e Tecnol\'ogico - CNPq (contract 307833/2022-4), and Funda\c c\~ao de Amparo \`a Pesquisa do Estado de S\~ao Paulo - FAPESP (contract 2022/00808-7). LRL is supported by the grant 2022/15107-4, São Paulo Research Foundation (FAPESP).

\appendix

\section{The proof of the relation \rf{fine}}
\label{sec:appendix2}
\setcounter{equation}{0}

Let us introduce the non-negative real-valued function $g$ as
\be  g\(\zeta\)\equiv \frac{4\,\sin^2 \(\frac{f}{2}\)}{f^{\prime 2}\,\zeta^2} =\frac{\(a^2+\zeta^2\)^2}{4\,m^2\,a^2\,\zeta^2}\sin^2 b  ;\qquad\qquad b\equiv 2\,m\,\arctan\(\(\frac{a}{\zeta}\)^{\eta'}\) \lab{fine2}\ee
and let us proof that
\be g \leq 1 \lab{maing}\ee
The function $g$ and its first-order derivative are continuous and $g$ satisfies $g(0)=g(\infty)=1$. Therefore, the maximum of $g$ must be at a critical point $\zeta_p$ or at $\zeta=0,\,1$. At $\zeta=a$ we have $g(a)=m^{-2}\,\sin^2\(m\,\frac{\pi}{2}\)\leq 1$. Clearly, for $m=1$ we have $g=1$, which satisfies \rf{maing}. From now on, we will study $g(\zeta)$ for $m\geq 2$ and for $\zeta$ lying on the interval $I\equiv (0,\,\infty)/\{a\}$. 
The critical points $\zeta_c$ in the interval $I$ corresponds to the solutions of
\be \sin^2 b_c = \eta\,\frac{2\,m\,a\,\zeta}{\(\zeta^2-a^2\)}\,\sin\(b_c\)\,\cos\(b_c\) \lab{relsin}\ee
where $b_c\equiv b|_{\zeta=\zeta_c}$. We can break the solutions of the equation \rf{relsin} in two types, corresponding to those cases where $\sin b_c =0$ and $\sin b_c \neq 0$. Clearly, if the critical point satisfies $\sin b_c =0$, which solves automatically \rf{relsin}, we have $g(\zeta_c)=0$. Otherwise, the equation \rf{relsin} is reduced to $\sin b_c = d\,\cos\(b_c\)$, with $d\equiv\frac{2\,m\,a\,\zeta}{\(\zeta^2-a^2\)}$, which leads to
\be b_c = \arctan(d) + \pi\,n_c \,;\qquad\qquad\qquad \forall n_c \in \IZ \ee
For this case we have $\sin^2 b_c=\frac{d^2}{1+d^2}$, which reduces \rf{fine2} to
\be
g_c=  \left[1+4\,\(m^2-1\)\,\(\frac{\(\frac{\zeta}{a}\)}{\(1+\frac{\zeta}{a}\)^2}\)^2\right]^{-1} \leq 1
\ee
Therefore, in the interval $I$ with $m\geq 2$ we have $g(\zeta)\leq 1$, completing the proof of \rf{maing}.


\begin{thebibliography}{10}

\bibitem{laf2017}
L.~A. Ferreira, ``{Exact self-duality in a modified Skyrme model},''
  \href{https://dx.doi.org/10.1007/JHEP07(2017)039}{{\em JHEP} {\bfseries 07}
  (2017) 039}, \href{https://arxiv.org/abs/1705.01824}{{\ttfamily
  arXiv:1705.01824 [hep-th]}}.

\bibitem{Ferreira:2024ivq}
L.~A. Ferreira and L.~R. Livramento, ``{Harmonic, Holomorphic and Rational Maps
  from Self-Duality},'' \href{https://arxiv.org/abs/2412.02636}{{\ttfamily
  arXiv:2412.02636 [hep-th]}}.

\bibitem{Ioannidou:1998zg}
T.~A. Ioannidou, B.~Piette, and W.~J. Zakrzewski, ``{Low-energy states in the
  SU(N) Skyrme models},'' in {\em {International Meeting on Mathematical
  Methods in Modern Theoretical Physics (ISPM 98)}}, pp.~91--123.
\newblock 11, 1998.
\newblock \href{https://arxiv.org/abs/hep-th/9811071}{{\ttfamily
  arXiv:hep-th/9811071}}.

\bibitem{harmonicmaps}
J.~Eells and L.~Lemaire, {\em Two reports on harmonic maps}.
\newblock World Scientific Publishing Company, 1995.

\bibitem{genbps}
C.~Adam, L.~A. Ferreira, E.~da~Hora, A.~Wereszczynski, and W.~J. Zakrzewski,
  ``{Some aspects of self-duality and generalised BPS theories},''
  \href{https://dx.doi.org/10.1007/JHEP08(2013)062}{{\em JHEP} {\bfseries 08}
  (2013) 062}, \href{https://arxiv.org/abs/1305.7239}{{\ttfamily
  arXiv:1305.7239 [hep-th]}}.

\bibitem{rajaraman1982solitons}
R.~Rajaraman, {\em Solitons and Instantons: An Introduction to Solitons and
  Instantons in Quantum Field Theory}.
\newblock North-Holland personal library.

\bibitem{mantonbook}
N.~S. Manton and P.~Sutcliffe,
  \href{https://dx.doi.org/10.1017/CBO9780511617034}{{\em {Topological
  solitons}}}.
\newblock Cambridge Monographs on Mathematical Physics. Cambridge University
  Press, 2004.

\bibitem{Shnir:book1}
Y.~M. Shnir, {\em {Topological and Non-Topological Solitons in Scalar Field
  Theories}}.
\newblock Cambridge University Press, 7, 2018.

\bibitem{Jackiw:1990aw}
R.~Jackiw and E.~J. Weinberg, ``{Self-dual Chern-Simons Vortices},''
  \href{https://dx.doi.org/10.1103/PhysRevLett.64.2234}{{\em Phys. Rev. Lett.}
  {\bfseries 64} (1990) 2234}.

\bibitem{luiz1}
L.~A. Ferreira and L.~R. Livramento, ``{Self-Duality in the Context of the
  Skyrme Model},'' \href{https://dx.doi.org/10.1007/JHEP09(2020)031}{{\em JHEP}
  {\bfseries 09} (2020) 031},
  \href{https://arxiv.org/abs/2004.08295}{{\ttfamily arXiv:2004.08295
  [hep-th]}}.

\bibitem{Ferreira:2021uhk}
L.~A. Ferreira and H.~Malavazzi, ``{Generalized self-duality for the
  Yang-Mills-Higgs system},''
  \href{https://dx.doi.org/10.1103/PhysRevD.104.105016}{{\em Phys. Rev. D}
  {\bfseries 104} no.~10, (2021) 105016},
  \href{https://arxiv.org/abs/2106.16182}{{\ttfamily arXiv:2106.16182
  [hep-th]}}.

\bibitem{mantonruback}
N.~S. Manton and P.~J. Ruback, ``{Skyrmions in Flat Space and Curved Space},''
  \href{https://dx.doi.org/10.1016/0370-2693(86)91271-2}{{\em Phys. Lett. B}
  {\bfseries 181} (1986) 137--140}.

\bibitem{derek}
D.~Harland, ``{Topological energy bounds for the Skyrme and Faddeev models with
  massive pions},''
  \href{https://dx.doi.org/10.1016/j.physletb.2013.11.062}{{\em Phys. Lett. B}
  {\bfseries 728} (2014) 518--523},
  \href{https://arxiv.org/abs/1311.2403}{{\ttfamily arXiv:1311.2403 [hep-th]}}.

\bibitem{skyrme1}
T.~H.~R. Skyrme, ``{A Nonlinear field theory},''
  \href{https://dx.doi.org/10.1098/rspa.1961.0018}{{\em Proc. Roy. Soc. Lond.
  A} {\bfseries 260} (1961) 127--138}.

\bibitem{skyrme2}
T.~H.~R. Skyrme, ``{A Unified Field Theory of Mesons and Baryons},''
  \href{https://dx.doi.org/10.1016/0029-5582(62)90775-7}{{\em Nucl. Phys.}
  {\bfseries 31} (1962) 556--569}.

\bibitem{adkins}
G.~S. Adkins, C.~R. Nappi, and E.~Witten, ``{Static Properties of Nucleons in
  the Skyrme Model},''
  \href{https://dx.doi.org/10.1016/0550-3213(83)90559-X}{{\em Nucl. Phys. B}
  {\bfseries 228} (1983) 552}.

\bibitem{derrick}
G.~H. Derrick, ``{Comments on nonlinear wave equations as models for elementary
  particles},'' \href{https://dx.doi.org/10.1063/1.1704233}{{\em J. Math.
  Phys.} {\bfseries 5} (1964) 1252--1254}.

\bibitem{Callan:1983nx}
C.~G. Callan, Jr. and E.~Witten, ``{Monopole Catalysis of Skyrmion Decay},''
  \href{https://dx.doi.org/10.1016/0550-3213(84)90088-9}{{\em Nucl. Phys. B}
  {\bfseries 239} (1984) 161--176}.

\bibitem{Piette:1997ny}
B.~M. A.~G. Piette and D.~H. Tchrakian, ``{Static solutions in the U(1) gauged
  Skyrme model},'' \href{https://dx.doi.org/10.1103/PhysRevD.62.025020}{{\em
  Phys. Rev. D} {\bfseries 62} (2000) 025020},
  \href{https://arxiv.org/abs/hep-th/9709189}{{\ttfamily
  arXiv:hep-th/9709189}}.

\bibitem{Radu:2005jp}
E.~Radu and D.~H. Tchrakian, ``{Spinning U(1) gauged skyrmions},''
  \href{https://dx.doi.org/10.1016/j.physletb.2005.10.020}{{\em Phys. Lett. B}
  {\bfseries 632} (2006) 109--113},
  \href{https://arxiv.org/abs/hep-th/0509014}{{\ttfamily
  arXiv:hep-th/0509014}}.

\bibitem{Livramento:2023keg}
L.~R. Livramento, E.~Radu, and Y.~Shnir, ``{Solitons in the Gauged
  Skyrme-Maxwell Model},''
  \href{https://dx.doi.org/10.3842/SIGMA.2023.042}{{\em SIGMA} {\bfseries 19}
  (2023) 042}, \href{https://arxiv.org/abs/2301.12848}{{\ttfamily
  arXiv:2301.12848 [hep-th]}}.

\bibitem{Livramento:2023tmm}
L.~R. Livramento and Y.~Shnir, ``{Multisolitons in a gauged Skyrme-Maxwell
  model},'' \href{https://dx.doi.org/10.1103/PhysRevD.108.065010}{{\em Phys.
  Rev. D} {\bfseries 108} no.~6, (2023) 065010},
  \href{https://arxiv.org/abs/2307.05756}{{\ttfamily arXiv:2307.05756
  [hep-th]}}.

\bibitem{luiz:false}
L.~A. Ferreira and L.~R. Livramento, ``{A false vacuum Skyrme model for nuclear
  matter},'' \href{https://dx.doi.org/10.1088/1361-6471/ac9226}{{\em J. Phys.
  G} {\bfseries 49} no.~11, (2022) 115102},
  \href{https://arxiv.org/abs/2106.13335}{{\ttfamily arXiv:2106.13335
  [hep-th]}}.

\bibitem{quasi-self-dual}
L.~A. Ferreira and L.~R. Livramento, ``{Quasi-self-dual Skyrme model},''
  \href{https://dx.doi.org/10.1103/PhysRevD.106.045003}{{\em Phys. Rev. D}
  {\bfseries 106} no.~4, (2022) 045003},
  \href{https://arxiv.org/abs/2205.13002}{{\ttfamily arXiv:2205.13002
  [hep-th]}}.

\bibitem{coleman1}
S.~R. Coleman, ``{The Fate of the False Vacuum. 1. Semiclassical Theory},''
  \href{https://dx.doi.org/10.1103/PhysRevD.16.1248}{{\em Phys. Rev. D}
  {\bfseries 15} (1977) 2929--2936}. [Erratum: Coleman,S., Phys.Rev.D 16, 1248
  (1977)].

\bibitem{coleman2}
S.~R. Coleman, V.~Glaser, and A.~Martin, ``{Action Minima Among Solutions to a
  Class of Euclidean Scalar Field Equations},''
  \href{https://dx.doi.org/10.1007/BF01609421}{{\em Commun. Math. Phys.}
  {\bfseries 58} (1978) 211--221}.

\bibitem{colemannew}
C.~G. Callan, Jr. and S.~R. Coleman, ``{The Fate of the False Vacuum. 2. First
  Quantum Corrections},''
  \href{https://dx.doi.org/10.1103/PhysRevD.16.1762}{{\em Phys. Rev. D}
  {\bfseries 16} (1977) 1762--1768}.

\bibitem{rational1}
``Rational maps, monopoles and skyrmions,'' {\em Nuclear Physics B} {\bfseries
  510} no.~3, (1998) 507--537. DOI: 10.1016/S0550-3213(97)00619-6.

\bibitem{rational2}
R.~A. Battye and P.~M. Sutcliffe, ``{Skyrmions, fullerenes and rational
  maps},'' {\em Reviews in Mathematical Physics} {\bfseries 14} no.~1, (2002)
  29--86, \href{https://arxiv.org/abs/hep-th/0103026}{{\ttfamily
  arXiv:hep-th/0103026 [hep-th]}}.
DOI: 10.1142/S0129055X02001065.

\bibitem{witten1}
E.~Witten, ``{Global Aspects of Current Algebra},''
  \href{https://dx.doi.org/10.1016/0550-3213(83)90063-9}{{\em Nucl. Phys. B}
  {\bfseries 223} (1983) 422--432}.

\bibitem{Holzwarth:1985rb}
G.~Holzwarth and B.~Schwesinger, ``{Baryons in the Skyrme Model},''
  \href{https://dx.doi.org/10.1088/0034-4885/49/8/001}{{\em Rept. Prog. Phys.}
  {\bfseries 49} (1986) 825}.

\bibitem{Weigel:1995cz}
H.~Weigel, ``{Baryons as three flavor solitons},''
  \href{https://dx.doi.org/10.1142/S0217751X96001218}{{\em Int. J. Mod. Phys.
  A} {\bfseries 11} (1996) 2419--2544},
  \href{https://arxiv.org/abs/hep-ph/9509398}{{\ttfamily
  arXiv:hep-ph/9509398}}.

\bibitem{Schechter:1999hg}
J.~Schechter and H.~Weigel, ``{The Skyrme model for baryons},''
  \href{https://arxiv.org/abs/hep-ph/9907554}{{\ttfamily
  arXiv:hep-ph/9907554}}.

\bibitem{Loiseau:1989ja}
B.~Loiseau, ``{Skyrions and Effective Lagrangians},''
  \href{https://dx.doi.org/10.1139/y89-186}{{\em Can. J. Phys.} {\bfseries 67}
  (1989) 1168--1179}.

\bibitem{helgason}
S.~Helgason, {\em Differential geometry, Lie groups, and symmetric spaces}.
\newblock Academic Press, Inc.,New York, NY, 01, 1978.

\bibitem{rational0}
S.~K. Donaldson, ``{Nahm's equations and the classification of monopoles},''
  \href{https://dx.doi.org/10.1007/BF01214583}{{\em Commun. Math. Phys.}
  {\bfseries 96} (1984) 387--407}.

\end{thebibliography}

\providecommand{\href}[2]{#2}\begingroup\raggedright\endgroup

\end{document}